\documentclass[aos,MSNbibl,seceqn,dvips]{arximspdf}
\usepackage{algorithm}
\usepackage{graphicx}
\doi{10.1214/13-AOS1167} 
\volume{42}
\issue{1}
\pubyear{2014}
\firstpage{115}
\lastpage{141}

\makeatletter
\newcommand{\rrVert}{\Vert}
\newcommand{\rrvert}{\vert}
\newcommand{\llVert}{\Vert}
\newcommand{\llvert}{\vert}
\newproclaim{rem}{Remark}

\newtheorem{prop}{Proposition}
\newtheorem{lem}{Lemma}
\newproclaim{defn*}{Definition}
\newcounter{hypA}

\newcommand{\iid}{\stackrel{\mathrm{i.i.d.}}{\sim}}

\newcommand{\esssup}{\operatorname{ess\,sup}}
\makeatother

\begin{document}
\begin{frontmatter}

\title{Twisted particle filters}
\runtitle{Twisted particle filters}

\begin{aug}
\author[a]{\fnms{Nick} \snm{Whiteley}\corref{}\thanksref{t1}\ead[label=e1]{nick.whiteley@bristol.ac.uk}}
\and
\author[b]{\fnms{Anthony} \snm{Lee}\thanksref{t2}\ead[label=e2]{anthony.lee@warwick.ac.uk}}
\affiliation{University of Bristol and University of Warwick}
\thankstext{t1}{Supported by EPSRC Grant EP/K023330/1.}
\thankstext{t2}{Supported by CRiSM, an EPSRC-HEFCE UK grant.}
\address[a]{School of Mathematics\\
University of Bristol\\
University Walk\\
Bristol BS8 1TW\\
United Kingdom\\
\printead{e1}}
\address[b]{Department of Statistics\\
University of Warwick\\
Coventry CV4 7AL\\
United Kingdom\\
\printead{e2}}
\runauthor{N. Whiteley and A. Lee}
\end{aug}

\received{\smonth{10} \syear{2012}}
\revised{\smonth{5} \syear{2013}}

%
\begin{abstract}
We investigate sampling laws for particle algorithms and the influence
of these laws on the efficiency of particle approximations of marginal
likelihoods in hidden Markov models. Among a broad class of candidates
we characterize the essentially unique family of particle system transition
kernels which is optimal with respect to an asymptotic-in-time variance
growth rate criterion. The sampling structure of the algorithm defined
by these optimal transitions turns out to be only subtly different
from standard algorithms and yet the fluctuation properties of the
estimates it provides can be dramatically different. The structure
of the optimal transition suggests a new class of algorithms, which
we term ``twisted'' particle filters and which we validate with asymptotic
analysis of a more traditional nature, in the regime where the number
of particles tends to infinity.
\end{abstract}

%

\end{frontmatter}

\section{Introduction}\label{sec:Introduction}

A hidden Markov model (HMM) with measurable state space $ (\mathsf
{X},\mathcal{X} )$
and observation space $ (\mathsf{Y},\mathcal{Y} )$ is a
process $ \{  (X_{n},Y_{n} );n\geq0 \} $ where
$ \{ X_{n};n\geq0 \} $ is a Markov chain on $\mathsf{X}$,
and each observation $Y_{n}$, valued in $\mathsf{Y}$, is conditionally
independent of the rest of the process given $X_{n}$. Let $\mu_{0}$
and $f$ be respectively a probability distribution and a Markov kernel
on $ (\mathsf{X},\mathcal{X} )$, and let $g$ be a Markov
kernel acting from $ (\mathsf{X},\mathcal{X} )$ to $
(\mathsf{Y},\mathcal{Y} )$,
with $g(x,\cdot)$ admitting a strictly positive density, denoted
similarly by $g(x,y)$, with respect to some dominating $\sigma$-finite
measure. The HMM specified by $\mu_{0}$, $f$ and $g$ is
%
\begin{eqnarray}\label{eq:HMM}
 X_{0}\sim\mu_{0}(\cdot),\qquad  X_{n}\vert
\{X_{n-1}=x_{n-1}\}& \sim& f(x_{n-1},\cdot),\qquad n
\geq1,
\nonumber
\\[-8pt]
\\[-8pt]
\nonumber
 Y_{n}\vert \{ X_{n}=x_{n} \} &\sim&
g(x_{n},\cdot),\qquad n\geq 0.
\end{eqnarray}
In practice, one often seeks to fit a HMM to data $ \{
Y_{0},Y_{1},\ldots \} $.
This motivates computation of the marginal likelihood of $ \{
Y_{0},Y_{1},\ldots \} $
under the model (\ref{eq:HMM}). We consider methods for approximate
performance of this computation.

Let $\Omega:=\mathsf{Y}^{\mathbb{Z}}$ be the set of doubly infinite
sequences valued in $\mathsf{Y}$. For $\omega= \{ \omega(n) \}
_{n\in\mathbb{Z}}\in\Omega$
we shall write the coordinate projection $Y_{n}(\omega)=\omega(n)$
and take as a recursive definition of the \emph{prediction filters},
the sequence of distributions $ \{ \pi_{n}^{\omega};n\geq0 \} $
given by
%
\begin{eqnarray}\label{eq:filtering_recursion}
\pi_{0}^{\omega}&:=&\mu_{0},
\nonumber
\\[-8pt]
\\[-8pt]
\nonumber
\pi_{n}^{\omega} (A )&:=&\frac{\int_{\mathsf{X}}\pi
_{n-1}^{\omega} (dx )g (x,Y_{n-1}(\omega)
)f(x,A)}{\int_{\mathsf{X}}\pi_{n-1}^{\omega} (dx )g
(x,Y_{n-1}(\omega) )},\qquad A\in\mathcal{X}, n
\geq1.
\end{eqnarray}
We are centrally concerned with the sequence $ \{ Z_{n}^{\omega
};n\geq0 \} $
defined by
%
\begin{equation}
Z_{0}^{\omega}:=1,\qquad Z_{n}^{\omega}:=Z_{n-1}^{\omega}
\int_{\mathsf{X}}\pi_{n-1}^{\omega} (dx )g
\bigl(x,Y_{n-1}(\omega ) \bigr),\qquad n\geq1.\label{eq:Z_recusion}
\end{equation}
Due to the conditional independence structure of the HMM, $\pi
_{n}^{\omega}$
is the conditional distribution of $X_{n}$ given $Y_{0:n-1}(\omega)$;
and $Z_{n}^{\omega}$ is the marginal likelihood evaluated at the
point $Y_{0:n-1}(\omega)$. The simplest particle filter, known as
the ``bootstrap'' algorithm \cite{gordon1993novel}, is given below.
It yields an approximation, $Z_{n}^{\omega,N}$, of each $Z_{n}^{\omega}$.

\begin{algorithm}[b]
\caption{Bootstrap particle filter}
\label{alg:bootstrap}
For $n=0$,

\qquad Sample $ (\zeta_{0}^{i} )_{i=1}^{N}\iid\mu_{0}$,

\qquad Report $Z_{0}^{\omega,N}=1$.

For $n\geq1$,

\qquad Report $Z_{n}^{\omega,N}=Z_{n-1}^{\omega,N}\cdot{
\frac{1}{N}\sum_{j=1}^{N}g(\zeta_{n-1}^{j},Y_{n-1}(\omega))}$,

\qquad Sample $ (\zeta_{n}^{i} )_{i=1}^{N}\vert
(\zeta_{n-1}^{i} )_{i=1}^{N} \iid \frac{\sum_{j=1}^{N}g(\zeta_{n-1}^{j},Y_{n-1}(\omega))f(\zeta_{n-1}^{j},\cdot
)}{\sum_{j=1}^{N}g(\zeta_{n-1}^{j},Y_{n-1}(\omega))}$.
\end{algorithm}

Convergence properties of particle algorithms in the regime
$N\rightarrow\infty$
are well understood \cite{del1999central,smc:the:C04,smc:the:K05,smc:the:DM08}
and their stability properties have been expressed through finite-$N$
error bounds \cite{smc:the:DMG01,smc:the:CdMG11,whiteley2011}, time-uniform
convergence \cite
{legland2003robustification,le2004stability,smc:the:OR05,smc:the:vH09}
and control on $N\rightarrow\infty$ asymptotic variance expressions
\cite{del2001interacting,favetto2012asymptotic,whiteley2011,douc2012long}.
Our aim is to rigorously address comparative questions of how and
why one algorithm may outperform another, and how it is possible
to modify standard algorithms in order to improve performance. Our
study is formulated in a generic framework which accommodates standard
particle algorithms and novel extensions. As an introduction we discuss
some of our intentions and findings in the context of the bootstrap
particle filter as per Algorithm~\ref{alg:bootstrap}; more precise
statements are given later.

Writing $\mathsf{E}_{N}^{\omega}$ for expectation with respect to
the law of the bootstrap particle filter processing a fixed observation
sequence $\omega\in\Omega$, the well-known lack-of-bias property
(\cite{smc:theory:Dm04}, Proposition~7.4.1) reads
%
\begin{equation}
\mathsf{E}_{N}^{\omega} \bigl[Z_{n}^{\omega,N}
\bigr]=Z_{n}^{\omega
},\label{eq:basic_unbias}
\end{equation}
and holds for any $n\geq0$ and $N\geq1$. This property is desirable
because it allows particle filters to be used within ``pseudo-marginal''-type
algorithms (see \cite{Andrieu2009} and references therein), and plays
a role in explaining the validity of some compound Monte Carlo techniques
such as particle Markov chain Monte Carlo \cite{Andrieu2010}. The
accuracy of $Z_{n}^{\omega,N}$ influences the performance of such
schemes \cite{Andrieu2012}. We shall analyze novel particle algorithms
arising through changes of measure on the left-hand side of (\ref
{eq:basic_unbias}),
similarly enjoying lack-of-bias, and which could therefore be used
in lieu of more standard particle filters in compound Monte Carlo
algorithms. The resulting approximation of $Z_{n}^{\omega}$ will
be of the form
%
\begin{equation}
\widetilde{Z}_{n}^{\omega,N}:=Z_{n}^{\omega,N}\cdot
\prod_{p=1}^{n}\phi _{p}^{\omega,N},\label{eq:basic_Z_tilde}
\end{equation}
where $Z_{n}^{\omega,N}$ is exactly the same functional of the particles
as in Algorithm~\ref{alg:bootstrap} and $ \{ \phi_{n}^{\omega
,N};n\geq1 \} $
is a sequence of functionals chosen such that, if we write $\widetilde
{\mathsf{E}}_{N}^{\omega}$
for expectation under the (as yet unspecified) alternative sampling
law, then the lack of bias property is preserved:
%
\begin{equation}
\widetilde{\mathsf{E}}_{N}^{\omega} \bigl[\widetilde{Z}_{n}^{\omega
,N}
\bigr]=Z_{n}^{\omega}.\label{eq:basic_unbias_2}
\end{equation}

Our main objective is to identify ``good'' choices of alternative
sampling laws, possibly allowing the transitions of the particles
to depend on past and/or future observations. Our criterion for performance
arises from a study of the normalized second moment of $\widetilde
{Z}_{n}^{\omega,N}$,
in the regime where $N$ is fixed and $n\rightarrow\infty$, in an
$\omega$-pathwise fashion.

For now, let us still consider $\omega\in\Omega$ as fixed. Then under
the probability law corresponding to Algorithm \ref{alg:bootstrap},
the generations of the particle system, $\zeta_{0},\zeta_{1},\ldots$
with $\zeta_{n}:=(\zeta_{n}^{1},\ldots,\zeta_{n}^{N})$, form an $\mathsf
{X}^{N}$-valued
time-inhomogeneous Markov chain. Let $ \{ \mathbf{M}^{\omega};\omega
\in\Omega \} $
be the family of Markov kernels such that for each $\omega\in\Omega$,
$\mathbf{M}^{\omega}\dvtx \mathsf{X}^{N}\times\mathcal{X}^{\otimes
N}\rightarrow[0,1]$
is given by
%
\begin{equation}
\mathbf{M}^{\omega}(x,dz)=\prod_{i=1}^{N}
\frac{\sum_{j=1}^{N}g
(x^{j},Y_{0}(\omega) )f (x^{j},dz^{i} )}{\sum_{j=1}^{N}g (x^{j},Y_{0}(\omega) )},\label{eq:basic_M_bold}
\end{equation}
with $x=(x^{1},\ldots,x^{N})\in\mathsf{X}^{N}$ and $z=(z^{1},\ldots
,z^{N})\in\mathsf{X}^{N}$.
Let $\theta\dvtx \Omega\rightarrow\Omega$ be the shift operator, $(\theta
\omega)(n):=\omega(n+1),  n\in\mathbb{Z},\omega\in\Omega$,
so that, for example, $Y_{0}(\theta\omega)=Y_{1}(\omega)$. The $n$-fold
iterate of $\theta$ will be written $\theta^{n}$ with $\theta^{0}=\mathrm{Id}$.
It is then clear that the sampling steps of Algorithm~\ref{alg:bootstrap}
implement
%
\begin{equation}
\zeta_{0}\sim\mu_{0}^{\otimes N},\qquad \zeta_{n}|
\zeta_{n-1}\sim \mathbf{M}^{\theta^{n-1}\omega}(\zeta_{n-1},\cdot),\qquad n
\geq1.\label
{eq:basic_transitions}
\end{equation}

\textit{Variance growth rates}.
For a family of Markov kernels $ \{ \widetilde{\mathbf{M}}^{\omega
};\omega\in\Omega \} $
belonging to a broad class of candidates and which may depend on $\omega$
in a rather general fashion, but subject to $\mathbf{M}^{\omega}(x,\cdot
)\ll\widetilde{\mathbf{M}}^{\omega}(x,\cdot)$
and other regularity conditions, we shall consider sampling the particle
system according to
%
\begin{equation}
\zeta_{0}\sim\mu_{0}^{\otimes N}, \qquad\zeta_{n}|
\zeta_{n-1}\sim \widetilde{\mathbf{M}}^{\theta^{n-1}\omega}(
\zeta_{n-1},\cdot),\qquad  n\geq1,\label{eq:basic_M_tilde_sample}
\end{equation}
and simply setting
%
\begin{equation}
\phi_{n}^{\omega,N}:=\frac{d\mathbf{M}^{\theta^{n-1}\omega}(\zeta
_{n-1},\cdot)}{d\widetilde{\mathbf{M}}^{\theta^{n-1}\omega}(\zeta
_{n-1},\cdot)}(\zeta_{n}),\qquad n
\geq1.\label{eq:rn_phi_defn}
\end{equation}
Then letting $\widetilde{\mathsf{E}}_{N}^{\omega}$ denote expectation
under the Markov law (\ref{eq:basic_M_tilde_sample}), and with
$\widetilde{Z}_{n}^{\omega,N}$
as in (\ref{eq:basic_Z_tilde}), we of course achieve (\ref{eq:basic_unbias_2}).

Let $\Omega$ be endowed with the product $\sigma$-algebra $\mathcal
{F}=\mathcal{Y}^{\otimes\mathbb{Z}}$
and let $\mathbb{P}$ be a probability measure on $ (\Omega,\mathcal
{F} )$.
We stress that $\mathbb{P}$ is not necessarily a measure on observation
sequences derived from the particular HMM (\ref{eq:HMM}), nor indeed
any HMM. Under the assumption that $\theta$ is $\mathbb{P}$-preserving
and ergodic, and under certain other regularity conditions, application
of our first main result, Proposition~\ref{prop:variance_growth},
establishes, for any \emph{fixed} $N\geq1$,\vspace*{1pt} existence of a deterministic
constant $\Upsilon_{N}(\widetilde{\mathbf{M}})$, depending on
$\widetilde{\mathbf{M}}= \{ \widetilde{\mathbf{M}}^{\omega};\omega
\in\Omega \} $
such that
%
\begin{equation}\quad \label{eq:basic_upsilon}
\frac{1}{n}\log\frac{\widetilde{\mathsf{E}}_{N}^{\omega} [
(\widetilde{Z}_{n}^{\omega,N} )^{2} ]}{ (Z_{n}^{\omega
} )^{2}}\longrightarrow\Upsilon_{N}(
\widetilde{\mathbf{M}}) \qquad\mbox{as } n\longrightarrow\infty, \mbox{ for }
\mathbb{P}\mbox{-a.a. } \omega.
\end{equation}
It must be the case that $\Upsilon_{N}(\widetilde{\mathbf{M}})\geq0$,
because variance is nonnegative and the lack of bias property (\ref
{eq:basic_unbias_2})
holds. We shall see that typically $\Upsilon_{N}(\widetilde{\mathbf{M}})>0$.

\textit{Optimal sampling}.
Our second main result (Theorem~\ref{thm-variance}) identifies,
for any fixed $N\geq1$ and among the class of candidates,
the essentially unique choice of the family $ \{ \widetilde{\mathbf
{M}}^{\omega};\omega\in\Omega \} $
which achieves $\Upsilon_{N}(\widetilde{\mathbf{M}})=0$. It turns
out that this optimal choice arises from a particular form of re-weighting
applied to each transition kernel $\mathbf{M}^{\omega}$ and is defined
in terms of a family of functions $ \{ h^{\omega}\dvtx \mathsf
{X\rightarrow\mathbb{R}_{+};\omega\in\Omega} \} $
which are, in abstract terms, generalized eigenfunctions associated
with algebraic structures underlying the particle algorithm. In the
context of the bootstrap particle filter, $h^{\omega}$ has the following
interpretation. $\pi_{n}^{\theta^{-n}\omega}$ is the prediction filter
initialized at time $-n$ and run forward to time zero, giving a distribution
over $X_{0}$ conditional on $Y_{-n}(\omega),\ldots,Y_{-1}(\omega)$.
Then, letting $\Pi_{n}^{\omega}$ be the distribution over $X_{0}$
obtained by further conditioning on $Y_{0}(\omega),\ldots,Y_{n-1}(\omega)$,
$h^{\omega}$ arises as the pointwise limit:
\[
h^{\omega}(x)=\lim_{n\rightarrow\infty}\frac{d\Pi_{n}^{\omega}}{d\pi
_{n}^{\theta^{-n}\omega}}(x).
\]
Theorem~\ref{thm-variance} establishes that for any $N\geq1$, $\Upsilon
_{N}(\widetilde{\mathbf{M}})=0$
if and only if, for $\mathbb{P}$-almost all $\omega\in\Omega$ there
exists a set $A_{\omega}\in\mathcal{X}^{\otimes N}$ such that $A_{\omega}^{c}$
is null (with respect to an as yet unnamed measure) and for any $x\in
A_{\omega}$,
%
\begin{equation}
\widetilde{\mathbf{M}}^{\omega}(x,B)=\frac{\int_{B}\mathbf{M}^{\omega
}(x,dx^{\prime})\mathbf{h}^{\theta\omega}(x^{\prime})}{\int_{\mathsf
{X}^{N}}\mathbf{M}^{\omega}(x,dx^{\prime})\mathbf{h}^{\theta\omega
}(x^{\prime})}\qquad\mbox{for all }B
\in\mathcal{X}^{\otimes N},\label
{eq:basic_twisted}
\end{equation}
where
\[
\mathbf{h}^{\omega} \dvtx x=\bigl(x^{1},\ldots,x^{N}\bigr)
\in\mathsf {X}^{N}\longmapsto N^{-1}\sum
_{i=1}^{N}h^{\omega}\bigl(x^{i}\bigr)
\in \mathbb{R}_{+}.
\]

In the rare-event and large deviations literatures, the action of
re-weighting Markov kernels using nonnegative eigenfunctions is
generically referred to as ``twisting.'' Since in the present context
we are applying re-weighting to the transitions of the entire particle
system, we shall adopt this terminology and consider a class of algorithms
which we refer to as \emph{twisted particle filters}.

\textit{Twisted particle filters}.
The form of the optimal transition (\ref{eq:basic_twisted}), where
$\mathbf{M}^{\omega}$ is re-weighted by an additive, nonnegative
functional, leads us to consider a new class of particle algorithms.
Consider a family of functions $ \{ \psi^{\omega}\dvtx \mathsf
{X}\rightarrow\mathbb{R}_{+};\omega\in\Omega \} $
and let $ \{ \widetilde{\mathbf{M}}^{\omega};\omega\in\Omega
\} $
be defined by
%
\begin{eqnarray}
\widetilde{\mathbf{M}}^{\omega}\bigl(x,dx^{\prime}\bigr)&=&
\frac{\mathbf{M}^{\omega
}(x,dx^{\prime})\bolds{\psi}^{\theta\omega}(x^{\prime})}
{\int_{\mathsf{X}^{N}}\mathbf{M}^{\omega}(x,dz)\bolds{\psi}^{\theta
\omega}(z)},\label{eq:basic_twisted-psi}
\\
\bolds{\psi}^{\omega} \dvtx x&=&\bigl(x^{1},\ldots,x^{N}
\bigr)\in\mathsf {X}^{N} \longmapsto N^{-1}\sum
_{i=1}^{N}\psi^{\omega}\bigl(x^{i}
\bigr) \in \mathbb{R}_{+}.\label{eq:basic_psi}
\end{eqnarray}
This setup clearly admits the optimal transition ($\psi^{\omega
}=h^{\omega}$)
and the standard transition (take $\psi^{\omega}=c$, for some positive
constant $c$) as special cases. Then introducing
\[
\widetilde{g}^{\omega}(x):=g \bigl(x,Y_{0}(\omega) \bigr)\int
_{\mathsf
{X}}f(x,dz)\psi^{\theta\omega}(z),
\]
we observe that $\phi_{n}^{\omega,N}$, defined in (\ref{eq:rn_phi_defn}),
is given by
%
\begin{equation}
\phi_{n}^{\omega,N}= \biggl[\frac{1}{N^{-1}\sum_{i=1}^{N}g(\zeta
_{n-1}^{i},Y_{n-1}(\omega))} \biggr]
\frac{\sum_{i=1}^{N}\widetilde
{g}^{\theta^{n-1}\omega}(\zeta_{n-1}^{i})}{\sum_{i=1}^{N}\psi^{\theta
^{n}\omega}(\zeta_{n}^{i})}.\label{eq:intro_rn_phi}
\end{equation}
Since $\bolds{\psi}^{\omega}$ is an additive functional, it
is clear that $\widetilde{\mathbf{M}}^{\omega}$ as per (\ref
{eq:basic_twisted-psi})
is of mixture form, and introducing the $\omega$-dependent Markov
kernel
\[
\widetilde{f}^{\omega}\bigl(x,dx^{\prime}\bigr):=\frac{f(x,dx^{\prime})\psi
^{\theta\omega}(x^{\prime})}{\int_{\mathsf{X}}f(x,dz)\psi^{\theta\omega}(z)},
\]
the procedure of sampling from (\ref{eq:basic_M_tilde_sample}) and
evaluating $\widetilde{Z}_{n}^{\omega,N}$ can be implemented through
Algorithm~\ref{alg:bootstrap_twisted}, in which $K_{n}$ and $A_{n}$
are auxiliary random variables employed for algorithmic purposes,
and the recursion for $\widetilde{Z}_{n}^{\omega,N}$ arises from
the definition of $Z_{n}^{\omega,N}$ combined with (\ref{eq:basic_Z_tilde})
and (\ref{eq:intro_rn_phi}).

\begin{algorithm}[t]
\caption{Twisted bootstrap particle filter}
\label{alg:bootstrap_twisted}
For $n=0$,

\qquad Sample $ (\zeta_{0}^{i} )_{i=1}^{N}\iid\mu_{0}$,

\qquad Report $\widetilde{Z}_{0}^{\omega,N}=1$.

For $n\geq1$,

\qquad Sample $K_{n}$ from the uniform distribution on $ \{
1,\ldots,N \} $,

\qquad Sample $A_{n}$ from the distribution on $ \{ 1,\ldots
,N \} $
with probabilities proportional

\qquad to
\[
\bigl\{ \widetilde{g}^{\theta^{n-1}\omega}\bigl(\zeta_{n-1}^{1}
\bigr),\ldots,\widetilde{g}^{\theta^{n-1}\omega}\bigl(\zeta_{n-1}^{N}
\bigr) \bigr\},
\]

\qquad Sample $\zeta_{n}^{K_{n}}\vert  \{ A_{n},K_{n},
(\zeta_{n-1}^{i} )_{i=1}^{N} \} \sim\widetilde{f}^{\theta
^{n}\omega}(\zeta_{n-1}^{A_{n}},\cdot)$,

\qquad Sample $ (\zeta_{n}^{i} )_{i\neq K_{n}}\vert  \{ K_{n}, (\zeta_{n-1}^{i} )_{i=1}^{N} \}  \iid
\frac{\sum_{j=1}^{N}g(\zeta_{n-1}^{j},Y_{n-1}(\omega))f (\zeta
_{n-1}^{j},\cdot )}{\sum_{j=1}^{N}g(\zeta_{n-1}^{j},Y_{n-1}(\omega))}$,

\qquad Report $\widetilde{Z}_{n}^{\omega,N}=\widetilde{Z}_{n-1}^{\omega
,N}\cdot{ \frac{\sum_{i=1}^{N}\widetilde{g}^{\theta
^{n-1}\omega}(\zeta_{n-1}^{i})}{\sum_{i=1}^{N}\psi^{\theta^{n}\omega
}(\zeta_{n}^{i})}}$.
\end{algorithm}

The difference between the sampling steps of Algorithm~\ref
{alg:bootstrap_twisted}
and Algorithm~\ref{alg:bootstrap} is fairly subtle: loosely speaking,
at each time step, $N-1$ of the particles in Algorithm~\ref
{alg:bootstrap_twisted}
are propagated by the same mechanism as in Algorithm~\ref{alg:bootstrap}.
However, with an appropriate choice of $\psi^{\omega}$, the fluctuation
properties of $\widetilde{Z}_{n}^{\omega,N}$under (\ref
{eq:basic_twisted-psi})--(\ref{eq:basic_psi})
can be dramatically different to those of $Z_{n}^{\omega,N}$ under
(\ref{eq:basic_transitions})--(\ref{eq:basic_M_bold}). Our third
main result (Theorem~\ref{thm:CLT}) concerns asymptotic fluctuation
properties of twisted particle approximations when $n$ and $\omega$
are fixed and $N\rightarrow\infty$. Under mild regularity conditions,
we prove central limit theorems for generic particle systems under
transitions like (\ref{eq:basic_twisted-psi})--(\ref{eq:basic_psi}).
For bounded functions $\varphi$ centered w.r.t. $\pi_{n}^{\omega}$,
we find that the $N\rightarrow\infty$ asymptotic variance associated
with $N^{-1/2}\sum_{i=1}^{N}\varphi(\zeta_{n}^{i})$ is the same when
sampled under Algorithms~\ref{alg:bootstrap} and~\ref{alg:bootstrap_twisted},
but the asymptotic variances of $\sqrt{N} (Z_{n}^{\omega
,N}-Z_{n}^{\omega} )$
and $\sqrt{N} (\widetilde{Z}_{n}^{\omega,N}-Z_{n}^{\omega} )$
are, in general, different.

The finite-$N$, finite-$n$ behavior of the relative variance of
the standard estimate $Z_{n}^{\omega,N}$ from Algorithm~\ref{alg:bootstrap}
is well understood. Under certain regularity assumptions, it can be
deduced from \cite{smc:the:CdMG11}, Theorem~5.1, that in our setting
$\Upsilon_{N}(\mathbf{M})$ must satisfy
%
\begin{equation}
\Upsilon_{N}(\mathbf{M})\leq\log \biggl[1+\frac{C}{N-1}
\biggr]\label
{eq:basic_upsilon _bound_boot}
\end{equation}
for some finite constant $C$ which depends on $g$ and $f$. Our
fourth main result (Proposition~\ref{Prop:upsilon-bound}) generalizes
(\ref{eq:basic_upsilon _bound_boot}) to the case of twisted particle
filters. With $\Upsilon_{N}(\widetilde{\mathbf{M}})$ as in (\ref
{eq:basic_upsilon}),
$\widetilde{\mathbf{M}}^{\omega}$ as in (\ref{eq:basic_twisted-psi}),
and under some regularity conditions,
\[
\Upsilon_{N}(\widetilde{\mathbf{M}})\leq\log \biggl[1+
\frac{C^{\prime
}}{N-1}\sup_{\omega,x,x^{\prime}}\biggl\llvert \frac{h^{\omega}(x)}{\psi^{\omega
}(x)}-
\frac{h^{\omega}(x^{\prime})}{\psi^{\omega}(x^{\prime})}\biggr\rrvert \biggr],
\]
where $C^{\prime}$ is a constant. Thus, whenever $\Upsilon_{N}(\mathbf{M})>0$,
by choosing $\psi$ ``close'' to $h$, we can in principle achieve
$\Upsilon_{N}(\widetilde{\mathbf{M}})<\Upsilon_{N}(\mathbf{M})$.

The rest of the paper is structured as follows. Section~\ref{sec:Nonnegative-kernels,-sampling}
introduces our general setting, addressing the generalized eigenvalue
properties of families of nonnegative kernels and sampling laws of
the particle systems we consider. Section~\ref{sec:Twisted-particle-algorithms}
narrows attention to twisted particle filters and considers some
properties in the regime where $N$ is fixed and $n\rightarrow\infty$,
and vice-versa. Section~\ref{sec:Discussion} discusses the application
of our main results to sequential importance sampling, bootstrap and
auxiliary particle filters. The proofs of Lemmas~\ref{Lem:degenerate}--\ref{lem:as-convergence},
Propositions~\ref{Prop:basic-log-like}--\ref{Prop:upsilon-bound}
and Theorems~\ref{thm-variance}--\ref{thm:CLT} are housed in the supplementary material \cite{tpf-supp}.

\section{Nonnegative kernels, sampling particles and variance
growth}
\label{sec:Nonnegative-kernels,-sampling}

\subsection{Notation and conventions}

Let $ (\mathsf{X},\mathcal{X} )$, $ (\mathsf{Y},\mathcal
{Y} )$,
$\Omega:=\mathsf{Y}^{\mathbb{Z}}$, $\mathcal{F}:=\mathcal{Y}^{\otimes
\mathbb{Z}}$,
and $\mathbb{P}$ and $\theta$ be as in Section~\ref{sec:Introduction}.
Expectation w.r.t. $\mathbb{P}$ will be denoted by $\mathbb{E}$.
Let $\mathcal{M}(\mathsf{X})$, $\mathcal{P}(\mathsf{X})$ and $\mathcal
{L}(\mathsf{X})$
be respectively the collections of measures, probability measures
and real-valued, bounded, $\mathcal{X}$-measurable functions on $\mathsf{X}$.
We write
\[
\llVert \varphi\rrVert:=\sup_{x}\bigl\llvert \varphi(x)
\bigr\rrvert
\]
and
%
\begin{equation}
\mu(\varphi):=\int_{\mathsf{X}}\varphi(x)\mu(dx) \qquad\mbox{for any }
\varphi\in\mathcal{L}(\mathsf{X}), \mu\in\mathcal{M}(\mathsf {X}).\label{eq:mu(phi)_notation}
\end{equation}

We will be dealing throughout with various real-valued functions on
$\Omega\times\mathsf{X}$ (and more generally $\Omega\times\mathsf{X}^{N}$,
etc.). For any such function $\varphi$, we write the $\omega$-section
of $\varphi$ as $\varphi^{\omega}\dvtx \mathsf{X}\rightarrow\mathbb{R}$,
$\varphi^{\omega}(x):=\varphi(\omega,x)$. For a function $\xi\dvtx \Omega
\rightarrow\mathbb{R}$
it will sometimes be convenient to write $\xi_{\omega}$ instead of
the more standard $\xi(\omega)$. We will need to express various
integration operations involving functions on $\Omega\times\mathsf{X}^{N}$
and their $\omega$-sections, so for completeness we quote the following
facts of measure theory (see, e.g., \cite{doob1994measure}, Chapter VI,
which will be used repeatedly without further comment): when $\varphi
\dvtx \Omega\times\mathsf{X}^{N}\rightarrow\mathbb{R}$
is measurable w.r.t. to $\mathcal{F}\otimes\mathcal{X}^{\otimes N}$,
then for every $\omega\in\Omega$, the $\omega$-section $\varphi^{\omega}$
is measurable w.r.t. $\mathcal{X}^{\otimes N}$; and, furthermore, for
any $\sigma$-finite measure $\mu$ on $ (\mathsf{X}^{N},\mathcal
{X}^{\otimes N})$,
if $\varphi$ is integrable w.r.t. to $\mathbb{P}\otimes\mu$, then
the function acting $\Omega\rightarrow\mathbb{R}$ which maps $\omega
\mapsto\mu(\varphi^{\omega})$
is measurable w.r.t. $\mathcal{F}$ and is $\mathbb{P}$-integrable.

Let $\varphi,\widetilde{\varphi}$ be two functions, each acting $\Omega
\times\mathsf{X}^{N}\rightarrow\mathbb{R}$
and each measurable w.r.t. $\mathcal{F}\otimes\mathcal{X}^{\otimes N}$.
We will need to talk about the sets on which such functions take the
same values. For any $\omega\in\Omega$, let $A_{\omega}:= \{ x\in
\mathsf{X}^{N}\dvtx \varphi^{\omega}(x)=\widetilde{\varphi}^{\omega}(x)
\} $
and let $\mu$ be a $\sigma$-finite measure on $ (\mathsf
{X}^{N},\mathcal{X}^{\otimes N} )$.
In order to avoid having to make the sets $ \{ A_{\omega};\omega\in
\Omega \} $
explicit in various statements, we will write by convention
\[
\mbox{\mbox{for }}\mathbb{P}\mbox{-a.a. } \omega,\qquad \varphi^{\omega
}(x)=\widetilde{
\varphi}^{\omega}(x)\qquad\mbox{\mbox{for }}\mu\mbox{-a.a. } x
\]
to mean $\mathbb{P} ( \{ \omega\dvtx \mu(A_{\omega}^{c})=0 \}
)=1$.

\subsection{Generalized eigenvalue theory for nonnegative kernels}
\label{sub:Generalized-eigen-value-theory}

Fix arbitrarily $\mu_{0}\in\mathcal{P}(\mathsf{X})$ and let $M\dvtx \mathsf
{\Omega}\times\mathsf{X}\times\mathcal{X}\rightarrow[0,1]$
be such that $M(\omega,x,\cdot)\in\mathcal{P}(\mathsf{X})$ for each
$(\omega,x)\in\Omega\times\mathsf{X}$, and $M(\cdot,\cdot,A)$ is
$\mathcal{F}\otimes\mathcal{X}$-measurable for each $A\in\mathcal{X}$.
Then for any $\omega$, $M(\omega,\cdot,\cdot)$ is a Markov kernel
on $(\mathsf{X},\mathcal{X})$ and when it is important to emphasize
this perspective, we shall often write $M^{\omega}(x,A)$ instead
of $M(\omega,x,A)$. We shall adopt similar notation for other kernels.

For any fixed $\omega\in\Omega$, let $\mathsf{E}^{\omega}$ denote
expectation with respect to the law of the time-inhomogeneous Markov
chain $ \{ X_{n};n\geq0 \} $, with each $X_{n}$ valued
in $\mathsf{X}$, initialized from $X_{0}\sim\mu_{0}$ and $X_{n}| \{
X_{n-1}=x_{n-1} \} \sim M^{\theta^{n-1}\omega}(x_{n-1},\cdot)$,
for $n\geq1$. Let $G\dvtx \Omega\times\mathsf{X}\rightarrow\mathbb{R}_{+}$
be a $\mathcal{F}\otimes\mathcal{X}$-measurable, strictly positive
and bounded function.
%
\begin{rem}
\label{rem:HMM}This setup is purposefully generic and accommodates,
as one particular instance, the case
%
\begin{equation}
G^{\omega}(x)=g\bigl(x,Y_{0}(\omega)\bigr), \qquad M^{\omega}
\bigl(x,dx^{\prime
}\bigr)=f\bigl(x,dx^{\prime}\bigr)\qquad \forall\omega\in
\Omega,\label{eq:hmm_as_FK}
\end{equation}
where $g$ and $f$ are as in Section~\ref{sec:Introduction}, and
then $\mathsf{E}^{\omega}[\prod_{p=0}^{n-1}g(X_{p},Y_{p}(\omega
))]=Z_{n}^{\omega}$
$=\mathsf{E}^{\omega}[\prod_{p=0}^{n-1}G^{\theta^{p}\omega}(X_{p})]$.
Other instances will be discussed in Section~\ref{sec:Discussion}.
\end{rem}
We next introduce two hypotheses. Since $\Omega:=\mathsf{Y}^{\mathbb{Z}}$,
({H1}) amounts to saying that the observation process is stationary
and ergodic. ({H2}) is a strong mixing condition that rarely
holds when $\mathsf{X}$ and $\mathsf{Y}$ are noncompact, and some
results do not rely on both (\ref{eq:H2_G}) and (\ref{eq:H2_M})
simultaneously but their combination allows us to avoid a layer of
technical presentation which would further lengthen and complicate
our proofs.

\begin{longlist}[(H2)]
\label{hyp:ergo}
\item[(H1)]The shift operator $\theta$ preserves $\mathbb{P}$ and is ergodic.
\item[(H2)]
\label{hyp:mixing}
There exist constants $\beta\in [1,\infty )$, $(\varepsilon
_{-},\varepsilon_{+})\in(0,\infty)^{2}$,
and $\nu\in\mathcal{P}(\mathsf{X})$ such that
%
\begin{eqnarray}
\frac{G(\omega,x)}{G(\omega^{\prime},x^{\prime})}&\leq&\beta\qquad \forall \bigl(\omega,\omega^{\prime},x,x^{\prime}
\bigr)\in\Omega^{2}\times \mathsf{X}^{2},\label{eq:H2_G}\\
\varepsilon_{-}\nu(\cdot)&\leq& M(\omega,x,\cdot)\leq\varepsilon_{+}
\nu(\cdot )\qquad \forall (\omega,x )\in\Omega\times\mathsf{X}.\label{eq:H2_M}
\end{eqnarray}
\end{longlist}

We now introduce the nonnegative kernel
%
\begin{equation}
Q\dvtx \mathsf{\Omega}\times\mathsf{X}\times\mathcal{X}\rightarrow\mathbb
{R}_{+},\qquad Q\bigl(\omega,x,dx^{\prime}\bigr):=G(\omega,x)M\bigl(
\omega,x,dx^{\prime}\bigr).\label{eq:Q_GM_defn}
\end{equation}
For any fixed $\omega\in\Omega$, we define the operators
%
\begin{eqnarray}
Q^{\omega}(\varphi) (x) &:= & \int_{\mathsf{X}}Q^{\omega}
\bigl(x,dx^{\prime
}\bigr)\varphi\bigl(x^{\prime}\bigr),\qquad \varphi\in
\mathcal{L}(\mathsf{X}),\label
{eq:Q_op_defn}
\\
\mu Q^{\omega}(\cdot) &:= & \int_{\mathsf{X}}
\mu(dx)Q^{\omega}(x,\cdot ), \qquad\mu\in\mathcal{M}(\mathsf{X}),\label{eq:Q_op_defn-1}
\end{eqnarray}
and let $ \{ Q_{n}^{\omega};n\in\mathbb{N} \} $ be defined
recursively by
%
\begin{equation}
Q_{0}^{\omega}:=\mathrm{Id}, \qquad Q_{n}^{\omega}=Q_{n-1}^{\omega}Q^{\theta
^{n-1}\omega},\qquad
n\geq1.\label{eq:Q_semigroup}
\end{equation}
This operator notation allows us to express
%
\begin{equation}
\mu_{0}Q_{n}^{\omega}(\varphi)=\mathsf{E}^{\omega}
\Biggl[\varphi (X_{n})\prod_{p=0}^{n-1}G^{\theta^{p}\omega}(X_{p})
\Biggr], \qquad n\geq 1,\varphi\in\mathcal{L}(\mathsf{X}).\label{eq:muQ_1=00003DE}
\end{equation}

It is well known that ({H1}) and ({H2}) together are
sufficient to establish the following result; see \cite{leroux1992maximum}
for related ideas in the context of HMMs.
%
\begin{prop}
\label{Prop:basic-log-like}Assume \textup{({H1})} and \textup{({H2})}.
Then there exists a constant $\Lambda\in(-\infty,\infty)$ independent
of the initial distribution $\mu_{0}\in\mathcal{P}(\mathsf{X})$ such
that
%
\begin{equation}
\frac{1}{n}\log\mathsf{E}^{\omega} \Biggl[\prod
_{p=0}^{n-1}G^{\theta
^{p}\omega}(X_{p}) \Biggr]
\rightarrow\Lambda\qquad \mbox{as } n\rightarrow \infty, \mathbb{P}\mbox{-a.s.}\label{eq:basic_log_like_conv}
\end{equation}
\end{prop}
It turns out that Proposition~\ref{Prop:basic-log-like} is one element
of a generalized eigenvalue theory for the nonnegative kernel $Q$.
Another element is Proposition~\ref{prop:eig}, which involves the
following objects. Let $\Phi^{\omega}\dvtx \mathcal{P}(\mathsf{X})\rightarrow
\mathcal{P}(\mathsf{X})$
be defined by
\[
\Phi^{\omega}(\mu)=\frac{\mu Q^{\omega}}{\mu Q^{\omega}(1)}, \qquad\mu\in \mathcal{P}(\mathsf{X}),
\]
and let $ \{ \Phi_{n}^{\omega};n\in\mathbb{N} \} $ be the
family of operators defined recursively by
\[
\Phi_{0}^{\omega}:=\mathrm{Id},\qquad \Phi_{n}^{\omega}:=
\Phi^{\theta^{n-1}\omega
}\circ\Phi_{n-1}^{\omega},
\]
so that each $\Phi_{n}^{\omega}$ acts $\mathcal{P}(\mathsf
{X})\rightarrow\mathcal{P}(\mathsf{X})$.
Under these definitions, for any $n\in\mathbb{N}$,
%
\begin{equation}
\Phi_{n}^{\omega}(\mu)=\frac{\mu Q_{n}^{\omega}}{\mu Q_{n}^{\omega
}(1)},\label{eq:Phi_write_R}
\end{equation}
which can be verified by induction, since from the above definitions
$\Phi_{0}^{\omega}=\mathrm{Id}$, $Q_{0}^{\omega}:=\mathrm{Id}$ and when (\ref{eq:Phi_write_R})
holds,
\[
\Phi_{n+1}^{\omega}(\mu)= \bigl(\Phi^{\theta^{n}\omega}\circ\Phi
_{n}^{\omega} \bigr) (\mu)=\frac{\Phi_{n}^{\omega}(\mu)Q^{\theta
^{n}\omega}}{\Phi_{n}^{\omega}(\mu)Q^{\theta^{n}\omega}(1)}=\frac{\mu
Q_{n}^{\omega}Q^{\theta^{n}\omega}}{\mu Q_{n}^{\omega}Q^{\theta
^{n}\omega}(1)}=
\frac{\mu Q_{n+1}^{\omega}}{\mu Q_{n+1}^{\omega}(1)}.
\]

\begin{rem}
In the setting $M^{\omega}(x,dx^{\prime}):=f(x,dx^{\prime})$, $G^{\omega
}(x):=g(x,Y_{0}(\omega))$,
then if $\mu_{0}$ and $\pi_{n}^{\omega}$ are respectively the initial
distribution and prediction-filter as in~(\ref{eq:filtering_recursion}),
we have
\[
\pi_{n+1}^{\omega}=\Phi^{\theta^{n}\omega} \bigl(
\pi_{n}^{\omega} \bigr),\qquad  n\geq0.
\]
\end{rem}

\begin{rem}
Under ({H2}), it is known that $\Phi_{n}^{\omega}$ is exponentially
stable with respect to initial conditions (e.g., \cite{smc:theory:Dm04}, Chapter~4)
that is, there exist constants $C<\infty$ and $\rho<1$ such that for
any $\varphi\in\mathcal{L}(\mathsf{X})$ and any $n\geq1$,
%
\begin{eqnarray}
\sup_{\omega\in\Omega}\sup_{\mu,\mu^{\prime}\in\mathcal{P}(\mathsf
{X})}\bigl\llvert \bigl[
\Phi_{n}^{\omega}(\mu)-\Phi_{n}^{\omega}\bigl(
\mu^{\prime
}\bigr) \bigr] (\varphi )\bigr\rrvert & \leq& \llVert \varphi
\rrVert C\rho^{n}.\label{eq:phi_conv_proof_1-1}
\end{eqnarray}
Equation~(\ref{eq:phi_conv_proof_1-1}) is used extensively in the proof of
the following proposition, which is a variation on the theme of Kifer's
Perron--Frobenius theorem for positive operators in a random environment
\cite{kifer1996perron}, Theorem~3.1.
\end{rem}
%
\begin{prop}
\label{prop:eig}Assume \textup{({H2})}.
\begin{longlist}[(1)]
\item[(1)] Fix $\mu\in\mathcal{P}(\mathsf{X})$. Then the limits
%
\begin{eqnarray}
\eta^{\omega}(A) &:= & \lim_{n\rightarrow\infty}\Phi_{n}^{\theta
^{-n}\omega}(
\mu) (A), \qquad\omega\in\Omega, A\in\mathcal{X},\label
{eq:pointwise_conv_1}
\\
h(\omega,x) &:= & \lim_{n\rightarrow\infty}\frac{Q_{n}^{\omega
}(1)(x)}{\Phi_{n}^{\theta^{-n}\omega}(\mu)Q_{n}^{\omega}(1)},\qquad \omega \in
\Omega, x\in\mathsf{X},\label{eq:pointwise_conv_2}
\end{eqnarray}
exist and define a family of probability measures $\eta:= \{ \eta
^{\omega}\in\mathcal{P}(\mathsf{X});\omega\in\Omega \} $
and an $\mathcal{F}\otimes\mathcal{X}$-measurable function $h\dvtx \Omega
\times\mathsf{X}\rightarrow\mathbb{R}$.

\item[(2)] In fact, $\eta$ and $h$ are independent of the particular $\mu$
chosen in part (1) and there exist constants $C<\infty$ and $\rho<1$
such that for any $\varphi\in\mathcal{L}(\mathsf{X})$,
%
\begin{equation}
\sup_{\omega\in\Omega}\sup_{\mu\in\mathcal{P}(\mathsf{X})}\bigl\llvert \bigl[
\Phi_{n}^{\theta^{-n}\omega}(\mu)-\eta^{\omega} \bigr](\varphi)\bigr
\rrvert \leq\llVert \varphi\rrVert C\rho^{n},\qquad n\geq1\label
{eq:eig_prop_lam}
\end{equation}
and
%
\begin{equation}
\sup_{\omega\in\Omega}\sup_{x\in\mathsf{X}}\sup
_{\mu\in\mathcal
{P}(\mathsf{X})}\biggl\llvert \frac{Q_{n}^{\omega}(1)(x)}{\Phi_{n}^{\theta
^{-n}\omega}(\mu)Q_{n}^{\omega}(1)}-h(\omega,x)\biggr\rrvert
\leq C\rho ^{n}, \qquad n\geq1.\label{eq:eig_prop_h}
\end{equation}
\item[(3)] $\lambda \dvtx \omega\in\Omega\longmapsto\eta^{\omega}(G^{\omega})\in
\mathbb{R}_{+}$
is measurable w.r.t. $\mathcal{F}$ and we have
%
\begin{equation}
\sup_{ (\omega,\omega^{\prime} )\in\Omega^{2}}\frac{\lambda
_{\omega}}{\lambda_{\omega^{\prime}}}<\infty,\qquad \sup
_{
(\omega,\omega^{\prime},x,x^{\prime} )\in\Omega^{2}\times\mathsf
{X}^{2}}\frac{h(\omega,x)}{h(\omega^{\prime},x^{\prime})}<\infty.\label
{eq:eig_prop_bounds}
\end{equation}
\item[(4)] Among all triples which consist of \textup{(i)} an $\Omega$-indexed
family of probability measures on $ (\mathsf{X},\mathcal{X} )$,
\textup{(ii)} an $\mathbb{R}_{+}$-valued, not identically zero, measurable
function on $\Omega\times\mathsf{X}$, and \textup{(iii)} a measurable function
on $\Omega$, the triple $ (\eta,h,\lambda )$, with $\eta,h$
as in part (1) and $\lambda$ as in part (3), uniquely satisfies the
system of equations
%
\begin{equation}\label{eq:eig_prop_eq}\qquad\quad
\eta^{\omega}Q^{\omega}=\lambda_{\omega}\eta^{\theta\omega},\quad
Q^{\omega}\bigl(h^{\theta\omega}\bigr)=\lambda_{\omega}h^{\omega},\quad
\eta ^{\omega}\bigl(h^{\omega}\bigr)=1\qquad
\mbox{for all } \omega\in \Omega.
\end{equation}
\end{longlist}
\end{prop}
The connection with Proposition~\ref{Prop:basic-log-like} is as follows:
%
\begin{prop}
\label{prop:Lam=00003Dlam}Assume \textup{({H1})}, \textup{({H2})}
and let $\Lambda$ be as in Proposition~\ref{Prop:basic-log-like}
and $\lambda$ be as in Proposition~\ref{prop:eig}. Then
%
\begin{equation}
\Lambda=\mathbb{E} [\log\lambda ]=\int_{\Omega}\log
\frac
{Q^{\omega}(h^{\theta\omega})(x)}{h^{\omega}(x)}\mathbb{P}(d\omega )\qquad\mbox{for any }x\in\mathsf{X}.\label{eq:Lambda_equal_expected_lambda}
\end{equation}
\end{prop}
In the setting of HMMs as per Remark~\ref{rem:HMM}, equalities like
the first one in (\ref{eq:Lambda_equal_expected_lambda}) appear routinely
in the study of likelihood-based estimators \cite
{leroux1992maximum,randal2011asymptotic}.
However, it is the second equality in (\ref{eq:Lambda_equal_expected_lambda}),
and generalizations thereof, which shall be crucial for our purposes
in the sequel.
%
\begin{rem}
If one weakens the ``$1$-step'' condition (\ref{eq:H2_M}) to an
$m$-step version for some $m\geq1$, then Propositions~\ref
{Prop:basic-log-like}--\ref{prop:Lam=00003Dlam}
can easily be generalized, working with the kernel $Q_{m}^{\omega}$
instead of $Q^{\omega}$. Part of the utility of the uniform in $\omega$
and $x$ bounds in ({H2}) is that various parts of Proposition~\ref{prop:eig} hold uniformly over $\omega\in\Omega$. If one allows
$\omega$-dependent constants and measures in (\ref{eq:H2_G}) and
(\ref{eq:H2_M}), and imposes certain explicit compactness and continuity
assumptions and ({H1}), then \cite{kifer1996perron},
Theorem~3.1,
provides a partial alternative to our Proposition~\ref{prop:eig}.
\end{rem}
We proceed by introducing the laws of the particle systems of interest.

\subsection{Law of the standard particle system}
\label{sub:The-standard-particle}

Unless stated otherwise, in this section we fix arbitrarily $N\geq1$
and write $\mathcal{P}(\mathsf{X}^{N})$ for the collection of probability
measures on $ (\mathsf{X}^{N},\mathcal{X}^{\otimes N} )$.

Let $\mathbf{M}\dvtx \mathsf{\Omega}\times\mathsf{X}^{N}\times\mathcal
{X}^{\otimes N}\rightarrow[0,1]$
be given, in integral form, by
%
\begin{equation}
\mathbf{M}(\omega,x,dz)=\prod_{i=1}^{N}
\biggl[\frac{\sum_{j=1}^{N}G(\omega,x^{j})M(\omega,x^{j},dz^{i})}{\sum_{j=1}^{N}G(\omega
,x^{j})} \biggr],\label{eq:standard_M_bold}
\end{equation}
where $x=(x^{1},\ldots,x^{N}),  z=(z^{1},\ldots,z^{N})\in\mathsf{X}^{N}$.
Each member of the family $ \{ \mathbf{M}^{\omega};\omega\in\Omega
\} $
is a Markov transition kernel for the entire $N$-particle\vadjust{\goodbreak} system
according to a ``multinomial'' resampling scheme with fitness function
$G(\omega,\cdot)$, followed by conditionally independent mutation
according to $M^{\omega}$.

Now for any given $\omega\in\Omega$, we shall denote by $\mathsf
{E}_{N}^{\omega}$
expectation with respect to the law of the Markov chain $ \{ \zeta
_{n};n\geq0 \} $,
with each $\zeta_{n}= \{ \zeta_{n}^{1},\ldots,\zeta_{n}^{N} \} $
valued in $\mathsf{X}^{N}$ and
%
\begin{equation}
\zeta_{0}\sim\mu_{0}^{\otimes N}, \qquad\zeta_{n}|
\zeta_{n-1}\sim \mathbf{M}^{\theta^{n-1}\omega}(\zeta_{n-1},
\cdot).\label{eq:basic_smc_law}
\end{equation}
We define, with $x=(x^{1},\ldots,x^{N})$,
%
\begin{equation}
\mathbf{G} \dvtx (\omega,x )\in\Omega\times\mathsf{X}^{N} \longmapsto
\frac{1}{N}\sum_{i=1}^{N}G\bigl(
\omega,x^{i}\bigr) \in \mathbb {R}_{+}.\label{eq:G_bold}
\end{equation}

\begin{rem}
\label{rem:unbias}For any $\varphi\in\mathcal{L}(\mathsf{X})$, if
we define the function
\[
\bolds{\varphi}\dvtx x=\bigl(x^{1},\ldots,x^{N}\bigr)\in
\mathsf{X}^{N}\longmapsto \frac{1}{N}\sum
_{i=1}^{N}\varphi\bigl(x^{i}\bigr)\in
\mathbb{R},
\]
then the lack-of-bias property of the particle approximation \cite{smc:theory:Dm04},
Proposition~7.4.1,
is
%
\begin{equation}
\mathsf{E}_{N}^{\omega} \Biggl[\bolds{\varphi}(
\zeta_{n})\prod_{p=0}^{n-1}
\mathbf{G}^{\theta^{p}\omega}(\zeta_{p}) \Biggr]=\mathsf {E}^{\omega}
\Biggl[\varphi(X_{n})\prod_{p=0}^{n-1}G^{\theta^{p}\omega
}(X_{p})
\Biggr].\label{eq:lack_of_bias}
\end{equation}
\end{rem}

\begin{rem}
When $M^{\omega}(x,\cdot)=f(x,\cdot)$ and $G^{\omega
}(x)=g(x,Y_{0}(\omega))$,
the sampling recipe for simulating the process $ \{ \zeta_{n};n\geq
0 \} $
according to (\ref{eq:basic_smc_law}) is the bootstrap particle filter:
Algorithm~\ref{alg:bootstrap}. Furthermore, the particle approximation
of $Z_{n}^{\omega}$ is then $\prod_{p=0}^{n-1}\mathbf{G}^{\theta
^{p}\omega}(\zeta_{p})$.
To see it is unbiased, apply (\ref{eq:lack_of_bias}) with $\varphi=1$.
\end{rem}

Part of our investigation will develop some limit theory for
%
\begin{equation}
\frac{\mathsf{E}_{N}^{\omega} [\prod_{p=0}^{n-1}\mathbf{G}^{\theta
^{p}\omega}(\zeta_{p})^{2} ]}{\mathsf{E}^{\omega} [\prod_{p=0}^{n-1}G^{\theta^{p}\omega}(X_{p}) ]^{2}},\label{eq:second_moment}
\end{equation}
when $N$ is fixed and $n\rightarrow\infty$. Our notation $\mathbf
{G},\mathbf{M}$
and (\ref{eq:lack_of_bias}) are intended to hint that the phenomena
described in Propositions~\ref{Prop:basic-log-like}--\ref{prop:Lam=00003Dlam}
are relevant to the study of~(\ref{eq:second_moment}). Indeed, this
is the direction in which we are heading. However, we will actually
study an object more general than (\ref{eq:second_moment}), arising
from a more general form of particle approximation, for the particle
system is distributed according to some Markov law, possibly different
to (\ref{eq:basic_smc_law}).

\subsection{Alternative sampling of the particle system}
\label{sub:Alternative-sampling-of}

Let us introduce $\widetilde{\mathbf{M}}\dvtx \mathsf{\Omega}\times\mathsf
{X}^{N}\times\mathcal{X}^{\otimes N}\rightarrow[0,1]$,
possibly different from $\mathbf{M}$. For fixed $\omega$, now denote
by $\mathbb{\widetilde{\mathsf{E}}}_{N}^{\omega}$ expectation with
respect to law of the Markov chain
%
\begin{equation}\label{eq:proposal_law}
\zeta_{0}\sim\mu_{0}^{\otimes N}, \qquad\zeta_{n}|
\zeta_{n-1}\sim \widetilde{\mathbf{M}}^{\theta^{n-1}\omega}(
\zeta_{n-1},\cdot).\vadjust{\goodbreak}
\end{equation}
We are going to specify a class of candidates for $\widetilde{\mathbf{M}}$,
and we first notice that the regularity condition {(H2)} transfers
to $\mathbf{G},\mathbf{M}$ in the following sense:
%
\begin{lem}
\label{Lem:bold_M_reg}Assume \textup{{(H2)}}. Then for any $N\geq1$,
\begin{eqnarray*}
\frac{\mathbf{G}(\omega,x)}{\mathbf{G}(\omega^{\prime},x^{\prime})}&\leq& \beta\qquad\forall \bigl(\omega,\omega^{\prime},x,x^{\prime}
\bigr)\in \Omega^{2}\times\mathsf{X}^{2N},
\\
\varepsilon_{-}^{N}\nu^{\otimes N}(\cdot)&\leq&\mathbf{M}(
\omega,x,\cdot)\leq \varepsilon_{+}^{N}\nu^{\otimes N}(
\cdot)\qquad \forall (\omega,x )\in\Omega\times\mathsf{X}^{N}.
\end{eqnarray*}
\end{lem}
The proof is omitted. We shall consider the following family of kernels.
\begin{defn*}[(of $\mathbb{M}$)]
Any $\widetilde{\mathbf{M}}\dvtx \mathsf{\Omega}\times\mathsf{X}^{N}\times
\mathcal{X}^{\otimes N}\rightarrow[0,1]$
is a member of $\mathbb{M}$ if and only if there exist constants
$ (\widetilde{\varepsilon}_{-},\widetilde{\varepsilon}_{+} )\in
(0,\infty )^{2}$
and $\widetilde{\nu}\in\mathcal{P} (\mathsf{X}^{N} )$ such
that
%
\begin{eqnarray}\label{eq:class_M_defn}
\widetilde{\nu} (\cdot )\widetilde{\varepsilon}_{-}&\leq& \widetilde{
\mathbf{M}}(\omega,x,\cdot)\leq\widetilde{\varepsilon }_{+}\widetilde{\nu}
(\cdot )\qquad \forall (\omega,x )\in\Omega\times\mathsf{X}^{N},
\nonumber
\\[-8pt]
\\[-8pt]
\nonumber
 \nu^{\otimes N}&\ll&\widetilde{\nu}\quad \mbox{and}\quad \int_{\mathsf
{X}^{N}}
\biggl(\frac{d\nu^{\otimes N}}{d\widetilde{\nu}} (x ) \biggr)^{2}\widetilde{\nu} (dx )<
\infty,
\end{eqnarray}
where $\nu$ is as in {(H2)}.
\end{defn*}

When $\widetilde{\mathbf{M}}$ is a member of $\mathbb{M}$ we write
%
\begin{equation}
\phi^{\omega}\bigl(x,x^{\prime}\bigr):=\frac{d\mathbf{M}^{\omega}(x,\cdot
)}{d\widetilde{\mathbf{M}}^{\omega}(x,\cdot)}
\bigl(x^{\prime}\bigr),\qquad \bigl(\omega,x,x^{\prime}\bigr)\in\Omega
\times\mathsf{X}^{2N},\label{eq:M_tilde_density}
\end{equation}
and in the context of sampling the particle system $ (\zeta
_{n};n\geq0 )$
under the law (\ref{eq:proposal_law}), we will take
%
\begin{equation}
\prod_{p=0}^{n-1}\mathbf{G}^{\theta^{p}\omega}(
\zeta_{p})\phi^{\theta
^{p}\omega}(\zeta_{p},
\zeta_{p+1})\label{eq:is_estimator}
\end{equation}
as an approximation of $\mathsf{E}^{\omega} [\prod_{p=0}^{n-1}G^{\theta^{p}\omega}(X_{p}) ]$.
In light of (\ref{eq:lack_of_bias}) and (\ref{eq:M_tilde_density}),
we have
\[
\widetilde{\mathsf{E}}_{N}^{\omega} \Biggl[\prod
_{p=0}^{n-1}\mathbf {G}^{\theta^{p}\omega}(
\zeta_{p})\phi^{\theta^{p}\omega}(\zeta_{p},\zeta
_{p+1}) \Biggr]=\mathsf{E}^{\omega} \Biggl[\prod
_{p=0}^{n-1}G^{\theta
^{p}\omega}(X_{p}) \Biggr].
\]
The following result describes the $n\rightarrow\infty$ behavior
of
%
\begin{equation}
\widetilde{\mathcal{V}}_{n,N}^{\omega}:=\frac{\widetilde{\mathsf
{E}}_{N}^{\omega} [\prod_{p=0}^{n-1}\mathbf{G}^{\theta^{p}\omega
}(\zeta_{p})^{2}\phi^{\theta^{p}\omega}(\zeta_{p},\zeta_{p+1})^{2}
]}{\mathsf{E}^{\omega} [\prod_{p=0}^{n-1}G^{\theta^{p}\omega
}(X_{p}) ]^{2}}.\label{eq:V_tilde_defn}
\end{equation}
Its proof starts by considering the family of kernels $ \{
\widetilde{\mathbf{R}}^{\omega};\omega\in\Omega \} $,
with
\[
\widetilde{\mathbf{R}}^{\omega}\bigl(x,dx^{\prime}\bigr):=
\mathbf{G}^{\omega
}(x)^{2}\phi^{\omega}\bigl(x,x^{\prime}
\bigr)^{2}\widetilde{\mathbf{M}}^{\omega
}\bigl(x,dx^{\prime}
\bigr),
\]
in terms of which the numerator of (\ref{eq:V_tilde_defn}) may be
written and which exhibit exactly similar properties to $Q^{\omega}$
appearing in the proof of Proposition~\ref{Prop:basic-log-like}.
%
\begin{prop}
\label{prop:variance_growth}Assume \textup{({H1})}, \textup{({H2})} and
fix $N\geq1$ arbitrarily. For every $\widetilde{\mathbf{M}}\in\mathbb{M}$
there exists a constant $\Upsilon_{N}(\widetilde{\mathbf{M}})\in
[0,\infty)$,
independent of the initial distribution $\mu_{0}$ such that
\[
\frac{1}{n}\log\widetilde{\mathcal{V}}_{n,N}^{\omega}
\longrightarrow \Upsilon_{N}(\widetilde{\mathbf{M}})\qquad \mbox{as } n
\rightarrow \infty, \mathbb{P}\mbox{-a.s.}
\]
\end{prop}
We now proceed to address the question of how $\Upsilon_{N}(\widetilde
{\mathbf{M}})$
depends on $\widetilde{\mathbf{M}}$. To this end, let us introduce
two further pieces of notation:
\[
\mathbf{Q}\bigl(\omega,x,dx^{\prime}\bigr):=\mathbf{G}(\omega,x)\mathbf{M}
\bigl(\omega,x,dx^{\prime}\bigr),
\]
and when {(H2) }holds, so that $h$ as in Proposition~\ref{prop:eig}
is well-defined, consider the function
%
\begin{equation}
\mathbf{h} \dvtx (\omega,x)\in\Omega\times\mathsf{X}^{N} \longmapsto
\frac{1}{N}\sum_{i=1}^{N}h\bigl(
\omega,x^{i}\bigr) \in \mathbb {R}_{+}.\label{eq:h_bold_defn}
\end{equation}
Our interest in (\ref{eq:h_bold_defn}) stems from the following pivotal
lemma, which shows how the generalized eigenfunction $h$ and eigenvalue
$\lambda$ of $Q$ appearing in Proposition~\ref{prop:eig} define
a generalized eigenfunction and eigenvalue for $\mathbf{Q}$, for
any $N\geq1$. Its proof is quite elementary, but is included here
for exposition since the structure it deals with underpins the algorithmic
developments in Section~\ref{sec:Twisted-particle-algorithms}.
%
\begin{lem}
\textup{\label{lem:eig_bold}For any $\omega\in\Omega$,
\[
\mathbf{Q}^{\omega} \bigl(\mathbf{h}^{\theta\omega} \bigr)=\lambda
_{\omega}\mathbf{h}^{\omega},
\]
where $\lambda_{\omega}$ is as in Proposition~\ref{prop:eig}.}
\end{lem}
\begin{pf}
\begin{eqnarray*}\hspace*{45pt}
\mathbf{Q}^{\omega} \bigl(\mathbf{h}^{\theta\omega} \bigr) (x) & = &
\frac
{1}{N}\sum_{k=1}^{N}\int
_{\mathsf{X}^{N}}\mathbf{Q}^{\omega
}(x,dz)h^{\theta\omega}
\bigl(z^{k}\bigr)
\\
& = & \frac{1}{N}\sum_{k=1}^{N}
\mathbf{G}^{\omega}(x)\int_{\mathsf
{X}}\frac{\sum_{i=1}^{N}Q^{\omega}(x^{i},dz^{k})}{\sum_{i=1}^{N}G^{\omega}(x^{i})}h^{\theta\omega}
\bigl(z^{k}\bigr)
\\
& = & \frac{1}{N}\sum_{k=1}^{N}
\frac{1}{N}\sum_{i=1}^{N}\int
_{\mathsf
{X}}Q^{\omega}\bigl(x^{i},dz^{k}
\bigr)h^{\theta\omega}\bigl(z^{k}\bigr)
\\
& = & \lambda_{\omega}\frac{1}{N}\sum_{i=1}^{N}h^{\omega
}
\bigl(x^{i}\bigr)=\lambda_{\omega}\mathbf{h}^{\omega}(x).\hspace*{105pt}\qed
\end{eqnarray*}
\noqed\end{pf}

Now consider taking
%
\begin{equation}
\widetilde{\mathbf{M}}^{\omega}\bigl(x,dx^{\prime}\bigr)=
\frac{\mathbf{M}^{\omega
}(x,dx^{\prime})\mathbf{h}^{\theta\omega}(x^{\prime})}{\int_{\mathsf
{X}^{N}}\mathbf{M}^{\omega}(x,dz)\mathbf{h}^{\theta\omega}(z)},\label
{eq:M_opt_preview}
\end{equation}
which is a member of $\mathbb{M}$, due to the definition of $\mathbf{h}$
and part (3) of Proposition~\ref{prop:eig}. In this case we have
%
\begin{eqnarray}\label{eq:opt_estimator}
&&\prod_{p=0}^{n-1}\mathbf{G}^{\theta^{p}\omega}(
\zeta_{p})\phi^{\theta
^{p}\omega}(\zeta_{p},
\zeta_{p+1})\nonumber\\[-2pt]
 & &\qquad=  \prod_{p=0}^{n-1}
\mathbf {G}^{\theta^{p}\omega}(\zeta_{p})\frac{\int_{\mathsf{X}^{N}}\mathbf
{M}^{\theta^{p}\omega}(\zeta_{p},dz_{p+1})\mathbf{h}^{\theta^{p+1}\omega
}(z_{p+1})}{\mathbf{h}^{\theta^{p+1}\omega}(\zeta_{p+1})}
\nonumber
\\[-2pt]
&&\qquad =  \prod_{p=0}^{n-1}\frac{\mathbf{Q}^{\theta^{p}\omega} (\mathbf
{h}^{\theta^{p+1}\omega} )(\zeta_{p})}{\mathbf{h}^{\theta
^{p+1}\omega}(\zeta_{p+1})}
\\[-2pt]
& &\qquad = \frac{\mathbf{h}^{\omega}(\zeta_{0})}{\mathbf{h}^{\theta
^{n}\omega}(\zeta_{n})}\prod_{p=0}^{n-1}
\frac{\mathbf{Q}^{\theta
^{p}\omega} (\mathbf{h}^{\theta^{p+1}\omega} )(\zeta
_{p})}{\mathbf{h}^{\theta^{p}\omega}(\zeta_{p})}
\nonumber
\\[-2pt]
&&\qquad =  \frac{\mathbf{h}^{\omega}(\zeta_{0})}{\mathbf{h}^{\theta
^{n}\omega}(\zeta_{n})}\prod_{p=0}^{n-1}
\lambda_{\theta^{p}\omega
},\nonumber
\end{eqnarray}
where the final equality is due to Lemma~\ref{lem:eig_bold}. Thus,
if we choose $\widetilde{\mathbf{M}}$ as per (\ref{eq:M_opt_preview}),
then the quantity in (\ref{eq:opt_estimator}) depends on the particle
system trajectory $\zeta_{0},\ldots,\zeta_{n}$ only through the quantities
$\mathbf{h}^{\omega}(\zeta_{0})$ and $\mathbf{h}^{\theta^{n}\omega
}(\zeta_{n})$,
and we then might hope that $\Upsilon_{N}(\widetilde{\mathbf{M}})=0$.
This turns out to be true, and much more strikingly, up to its definition
on certain sets of measure zero, $\widetilde{\mathbf{M}}$ as in (\ref
{eq:M_opt_preview})
is the unique member of $\mathbb{M}$ which achieves $\Upsilon
_{N}(\widetilde{\mathbf{M}})=0$,
in the sense of the following theorem.

\begin{thm}
\label{thm-variance}Assume \textup{({H1})}, \textup{({H2})}, let $N\geq1$
be fixed arbitrarily and assume $\widetilde{\mathbf{M}}$ belongs
to $\mathbb{M}$. Then \textup{(1)--(3)} are equivalent:
\begin{longlist}[(1)]
\item[(1)] $\Upsilon_{N}(\widetilde{\mathbf{M}})=0$.

\item[(2)] For $\mathbb{P}$-almost all $\omega\in\Omega$, there exists
$A_{\omega}\in\mathcal{X}^{\otimes N}$
such that $\nu^{\otimes N}(A_{\omega}^{c})=0$ and for any $x\in
A_{\omega}$,
%
\begin{equation}
\widetilde{\mathbf{M}}^{\omega}(x,B)=\frac{\int_{B}\mathbf{M}^{\omega
}(x,dx^{\prime})\mathbf{h}^{\theta\omega}(x^{\prime})}{\int_{\mathsf
{X}^{N}}\mathbf{M}^{\omega}(x,dz)\mathbf{h}^{\theta\omega}(z)}\qquad\mbox{for all }B
\in\mathcal{X}^{\otimes N}.\label{eq:special}
\end{equation}

\item[(3)] For $\mathbb{P}$-almost all $\omega\in\Omega$, $\sup_{n}\widetilde
{\mathcal{V}}_{n,N}^{\omega}<\infty$.
\end{longlist}
\end{thm}
The re-weighted particle transitions (\ref{eq:special}) are reminiscent
of certain eigenfunction transformations of general type
branching\vadjust{\goodbreak}
processes studied by Athreya \cite{athreya2000change} and more broadly
can be viewed as a randomized version of Doob's $h$-process.
See~\cite{tpf-supp} for further information. Time-homogeneous counterparts
of such transitions arise in the analysis of certain Markov chain
rare event problems \cite{bucklew1990monte}; in order to prove
$(1)\Rightarrow(2)$,
we generalize the proof of necessity in \cite{bucklew1990monte},
Theorem~3,
to the case of families of kernels driven by an ergodic, measure-preserving
transform.

The following lemma serves to accompany Proposition~\ref{prop:variance_growth}
and Theorem~\ref{thm-variance}, and provides necessary and sufficient
conditions for $\Upsilon_{N}(\widetilde{\mathbf{M}})=0$ in the case
of taking $\widetilde{\mathbf{M}}=\mathbf{M}$, that is, the transitions
of the standard particle system.
%
\begin{lem}
\label{Lem:degenerate}Assume \textup{({H1})}, \textup{({H2})} and let
$N\geq1$ be fixed arbitrarily. Then \textup{(1)--(3)} are equivalent:
\begin{longlist}[(1)]
\item[(1)] $\Upsilon_{N}(\mathbf{M})=0$.

\item[(2)] For $\mathbb{P}\mbox{-a.a. } \omega,  h^{\omega}(x)=1$,
for $\nu$\mbox{-a.a.} $x$.

\item[(3)] There exists a random variable $C\dvtx \Omega\rightarrow\mathbb{R}_{+}$
such that
\[
\mbox{for }\mathbb{P}\mbox{-a.a. } \omega,\qquad G^{\omega}(x)=C_{\omega}\qquad \mbox{for
}\nu\mbox{-a.a. } x.
\]
\end{longlist}
\end{lem}
In situations of practical interest, point (3) of Lemma~\ref{Lem:degenerate}
is usually false, and then it must be the case that $\Upsilon
_{N}(\mathbf{M})>0$.
It then appears that a choice of $\widetilde{\mathbf{M}}$ which approximates
the optimal transition, (\ref{eq:special}), may yield a provable
performance advantage over $\mathbf{M}$, in the sense of achieving
strict inequality $\Upsilon_{N}(\widetilde{\mathbf{M}})<\Upsilon
_{N}(\mathbf{M})$.
This leads us to consider the class of particle algorithms treated
in the next section.

\section{Twisted particle algorithms}
\label{sec:Twisted-particle-algorithms}

The form of the optimal transition kernel~(\ref{eq:special}) suggests
that we consider families of kernels arising from re-weighting of
$\mathbf{M}^{\omega}(x,\cdot)$ by an additive, nonnegative functional.
In this section we will analyze particle algorithms arising from kernels
of this general form. Let $\psi\dvtx \Omega\times\mathsf{X}\rightarrow\mathbb{R}_{+}$
be a strictly positive, bounded and measurable function and define
\[
\bolds{\psi} \dvtx (\omega,x )\in\Omega\times\mathsf {X}^{N}
\longmapsto \frac{1}{N}\sum_{i=1}^{N}
\psi\bigl(\omega,x^{i}\bigr) \in \mathbb{R}_{+}.
\]
For the purposes of this section, let us consider the following mild
regularity assumption:

\begin{longlist}[(H3)]
\item[(H3)]For each $\omega\in\Omega$, $\sup_{x}G^{\omega}(x)<\infty$ and $\sup_{x}\psi^{\omega}(x)<\infty$.
\end{longlist}

When {(H3)} holds the following Markov kernel is well-defined:
%
\begin{equation}
\widetilde{\mathbf{M}}^{\omega}\bigl(x,dx^{\prime}\bigr)=
\frac{\mathbf{M}^{\omega
}(x,dx^{\prime})\bolds{\psi}^{\theta\omega}(x^{\prime})}{\int_{\mathsf{X}^{N}}\mathbf{M}^{\omega}(x,dz)\bolds{\psi}^{\theta
\omega}(z)}.\label{eq:M_psi}
\end{equation}
We shall analyze particle approximations which arise from sampling
under (\ref{eq:M_psi}). Our motivation here is that we have in mind
situations where $\psi$ is chosen to be some approximation of $h$,
assuming the latter exists. The kernel (\ref{eq:M_psi}) accommodates
the standard transition (\ref{eq:standard_M_bold}) (e.g., take $\psi
^{\omega}=1$)
and the optimal transition identified in Theorem~\ref{thm-variance}
(take $\psi^{\omega}=h^{\omega}$). We note that (\ref{eq:M_psi})
depends on $\psi^{\omega}$only up to a constant of proportionality.

This section addresses two main objectives: First, to validate the
particle approximations delivered when sampling under (\ref{eq:M_psi}),
by analyzing some of their convergence and fluctuation properties
in the regime where $N\rightarrow\infty$.\vspace*{1pt} Second, to provide an
estimate of $\Upsilon_{N}(\widetilde{\mathbf{M}})$ which exhibits
dependence on $N$ and on the discrepancy between $\psi^{\omega}$
and $h^{\omega}$.

Let us introduce a little more notation. Define, for each $\omega\in
\Omega$,
the sequence of probability measures:
\[
\eta_{0}^{\omega}:=\mu_{0},\qquad \eta_{n}^{\omega}:=
\Phi^{\theta
^{n-1}\omega}\bigl(\eta_{n-1}^{\omega}\bigr),\qquad n\geq1.
\]

With $ \{ \zeta_{n};n\geq0 \} $ the sequence of generations
of the particles, we write
\begin{eqnarray*}
\eta_{n}^{N}&:=&\frac{1}{N}\sum
_{i=1}^{N}\delta_{\zeta_{n}^{i}}, \qquad n\geq0,
\\
\phi_{n}^{\omega,N}&:=&\frac{\eta_{n-1}^{N}Q^{\theta^{n-1}\omega}(\psi
^{\theta^{n}\omega})}{\eta_{n-1}^{N}(G^{\theta^{n-1}\omega})}\frac
{1}{\eta_{n}^{N}(\psi^{\theta^{n}\omega})}=
\frac{\Phi^{\theta
^{n-1}\omega} (\eta_{n-1}^{N} )(\psi^{\theta^{n}\omega})}{\eta
_{n}^{N}(\psi^{\theta^{n}\omega})},\qquad n\geq1,\\
\gamma_{0}^{\omega}&:=&\mu_{0},\qquad
\gamma_{n}^{\omega}(\varphi ):=\eta_{n}^{\omega}(
\varphi)\prod_{p=0}^{n-1}\eta_{p}^{\omega
}
\bigl(G^{\theta^{p}\omega}\bigr),\qquad n\geq1,
\\
\gamma_{0}^{\omega,N}&:=&\eta_{0}^{N}, \qquad
\gamma_{n}^{\omega
,N}(\varphi):=\eta_{n}^{N}(
\varphi)\prod_{p=0}^{n-1}\eta
_{p}^{N}\bigl(G^{\theta^{p}\omega}\bigr)\phi_{p+1}^{\omega,N},\qquad
n\geq1.
\end{eqnarray*}

To connect with (\ref{eq:M_tilde_density}), we note that for $\widetilde
{\mathbf{M}}$
as in (\ref{eq:M_psi}), we have
\begin{eqnarray*}
\phi_{n}^{\omega,N} & = & \Biggl(\frac{\mbox{1}}{N}\sum
_{i=1}^{N}G^{\theta^{n-1}\omega}\bigl(\zeta_{n-1}^{i}
\bigr) \Biggr)^{-1}\frac{\sum_{i=1}^{N}Q^{\theta^{n-1}\omega}(\psi^{\theta^{n}\omega})(\zeta
_{n-1}^{i})}{\sum_{i=1}^{N}\psi^{\theta^{n}\omega}(\zeta_{n}^{i})}
\\
& = & \phi^{\theta^{n}\omega}(\zeta_{n-1},\zeta_{n})
\end{eqnarray*}
and
\[
\prod_{p=0}^{n-1}\mathbf{G}^{\theta^{p}\omega}(
\zeta_{p})\phi^{\theta
^{p}\omega}(\zeta_{p},
\zeta_{p+1})=\prod_{p=0}^{n-1}
\frac{\sum_{i=1}^{N}Q^{\theta^{p}\omega}(\psi^{\theta^{p+1}\omega})(\zeta
_{p}^{i})}{\sum_{i=1}^{N}\psi^{\theta^{p+1}\omega}(\zeta
_{p+1}^{i})}=\gamma_{n}^{\omega,N}(1).
\]

Introducing the Markov kernel $\widetilde{M}^{\omega}(x,dx^{\prime
})\propto M^{\omega}(x,dx^{\prime})\psi^{\theta\omega}(x^{\prime})$,
Algorithm~\ref{alg:twisted} gives a recipe for sampling the particle
system according to (\ref{eq:proposal_law}) with $\widetilde{\mathbf{M}}$
as in~(\ref{eq:M_psi}) (details of the derivation of this algorithm
are given in~\cite{tpf-supp}). Here $K_{n}$ and $A_{n}$ are just
some auxiliary random variables introduced for algorithmic convenience.

\begin{algorithm}[t]
\caption{Twisted particle algorithm}
\label{alg:twisted}
For $n=0$,

\qquad Sample $ (\zeta_{0}^{i} )_{i=1}^{N}\iid\mu_{0}$
and report $\widetilde{Z}_{0}^{\omega,N}=1$.

For $n\geq1$,

\qquad Sample $K_{n}$ from the uniform distribution on $ \{
1,\ldots,N \} $,

\qquad Sample $A_{n}$ from the distribution on $ \{ 1,\ldots
,N \} $
with probabilities proportional

\qquad to
\[
\bigl\{ Q^{\theta^{n-1}\omega}\bigl(\psi^{\theta^{n}\omega}\bigr) \bigl(\zeta
_{n-1}^{1}\bigr),\ldots,Q^{\theta^{n-1}\omega}\bigl(
\psi^{\theta^{n}\omega}\bigr) \bigl(\zeta _{n-1}^{N}\bigr) \bigr\}
,
\]
%

\qquad Sample $\zeta_{n}^{K_{n}}\vert  \{ A_{n},K_{n},
(\zeta_{n-1}^{i} )_{i=1}^{N} \} \sim\widetilde{M}^{\theta
^{n}\omega}(\zeta_{n-1}^{A_{n}},\cdot)$,

\qquad Sample $ (\zeta_{n}^{i} )_{i\neq K_{n}}\vert  \{ K_{n}, (\zeta_{n-1}^{i} )_{i=1}^{N} \}  \iid
\frac{\sum_{j=1}^{N}G^{\theta^{n-1}\omega}(\zeta
_{n-1}^{j})M^{\theta^{n-1}\omega}(\zeta_{n-1}^{j},\cdot)}{\sum_{j=1}^{N}G^{\theta^{n-1}\omega}(\zeta_{n-1}^{j})}$,

\qquad Report $\widetilde{Z}_{n}^{\omega,N}=\widetilde{Z}_{n-1}^{\omega
,N}\cdot{ \frac{\sum_{i=1}^{N}Q^{\theta^{n-1}\omega}(\psi
^{\theta^{n}\omega})(\zeta_{n-1}^{i})}{\sum_{i=1}^{N}\psi^{\theta
^{n}\omega}(\zeta_{n}^{i})}}$.
\end{algorithm}

\subsection{\texorpdfstring{Analysis for $N\rightarrow\infty$}
{Analysis for N -> infinity}}

%
\begin{lem}
\label{lem:as-convergence}For each, $n\geq0$, fixed $\omega\in\Omega$
and $\varphi\in\mathcal{L}(\mathsf{X})$,
%
\begin{eqnarray}
\eta_{n}^{N}(\varphi)-\eta_{n}^{\omega}(
\varphi) & \longrightarrow& 0,\label{eq:as_convergence_eta}
\\
\gamma_{n}^{\omega,N}(\varphi)-\gamma_{n}^{\omega}(
\varphi) & \longrightarrow& 0\label{eq:as_convergence_gamma}
\end{eqnarray}
almost surely, as $N\rightarrow\infty$.
\end{lem}
Now define
\[
\overline{Q}_{n,n}^{\omega}:=\mathrm{Id},\qquad n\geq0,\qquad \overline
{Q}_{p,n}^{\omega}:=\frac{Q^{\theta^{p}\omega}\cdots Q^{\theta
^{n-1}\omega}}{\prod_{q=p}^{n-1}\eta_{q}^{\omega}(G^{\theta^{q}\omega
})},\qquad n\geq1, 0\leq p<n,
\]
and notice that $\mu_{0}\overline{Q}_{0,n}^{\omega}=\eta_{n}^{\omega}$.
%
\begin{thm}
\label{thm:CLT}Assume \textup{{(H3)}}. Then for any $n\geq0$, fixed
$\omega\in\Omega$ and $\varphi\in\mathcal{L}(\mathsf{X})$,
%
\begin{eqnarray}
\sqrt{N} \bigl[\gamma_{n}^{\omega,N}(\varphi)-\gamma_{n}^{\omega}(
\varphi ) \bigr] & \Rightarrow& \mathcal{N}\bigl(0,\varsigma_{n,\omega}^{2}(
\varphi )\bigr),\label{eq:clt_gamma}
\\
\sqrt{N} \bigl[\eta_{n}^{N}(\varphi)-\eta_{n}^{\omega}(
\varphi) \bigr] & \Rightarrow& \mathcal{N}\bigl(0,\sigma_{n,\omega}^{2}(
\varphi)\bigr)\label{eq:clt_eta}
\end{eqnarray}
as $N\rightarrow\infty$ where
%
\begin{equation}
\varsigma_{n,\omega}^{2}(\varphi)=\sum
_{p=0}^{n}\gamma_{p}^{\omega
}(1)^{2}
\eta_{p}^{\omega} \biggl[ \biggl(Q_{n-p}^{\theta^{p}\omega}(
\varphi )-\frac{\psi^{\theta^{p}\omega}}{\eta_{p}^{\omega} (\psi^{\theta
^{p}\omega} )}\eta_{p}^{\omega}Q_{n-p}^{\theta^{p}\omega}(
\varphi ) \biggr)^{2} \biggr],\label{eq:asymp_var_gamma}
\end{equation}
with the convention $\psi^{\omega}/\eta_{0}^{\omega}(\psi^{\omega})=1$,
and
%
\begin{equation}
\sigma_{n,\omega}^{2}(\varphi)=\sum
_{p=0}^{n}\eta_{p}^{\omega} \bigl[
\bigl(\overline{Q}_{p,n}^{\omega}\bigl(\varphi-
\eta_{n}^{\omega}(\varphi )\bigr) \bigr)^{2}
\bigr].\label{eq:asymp_var_eta}
\end{equation}
\end{thm}
%
\begin{rem}
The asymptotic variance expression (\ref{eq:asymp_var_eta}) is independent
of the particular choice of $\psi$ [the CLT holding subject to {(H3)},
of course] and is exactly the same expression obtained in the CLT
for the standard particle system (i.e., $\psi$ constant); see, for example,
\cite{smc:theory:Dm04}, Proposition~9.4.2. However, the asymptotic
variance in (\ref{eq:asymp_var_gamma}) clearly does depend on $\psi$
in general.
\end{rem}

\subsection{\texorpdfstring{Analysis for $n\rightarrow\infty$}
{Analysis for n -> infinity}}

For $\widetilde{\mathbf{M}}$ as in (\ref{eq:M_psi}), we obtain an
estimate of $\Upsilon_{N}(\widetilde{\mathbf{M}})$ which exhibits
its dependence on $N$ and the discrepancy between $\psi$ and $h$.
%
\begin{prop}
\label{Prop:upsilon-bound}Assume \textup{({H1})}, \textup{({H2})} and
$\sup_{\omega,\omega^{\prime}x,x^{`\prime}}\psi^{\omega}(x)/\psi^{\omega
^{\prime}}(x^{\prime})<\infty$.
Then for any $N\geq2$,
\[
\Upsilon_{N}(\widetilde{\mathbf{M}})\leq\log \biggl[1+
\frac
{1}{N-1}\mathcal{D}_{\mathbb{P}}(\psi,h) \biggr],
\]
where
\begin{eqnarray*}
\mathcal{D}_{\mathbb{P}}(\psi,h) &:= & \mathbb{P}-\esssup_{\omega}
\biggl\{ C_{\omega}\sup_{ (z,z^{\prime} )\in\mathsf{X}^{2}}\biggl\llvert
\frac{h^{\omega}(z)}{\psi^{\omega}(z)}-\frac{h^{\omega}(z^{\prime
})}{\psi^{\omega}(z^{\prime})}\biggr\rrvert \biggr\},
\\
C_{\omega} &:= & \biggl(2\sup_{z,z^{\prime}\in\mathsf{X}}\frac{\psi
^{\omega}(z)}{\psi^{\omega}(z^{\prime})}-1
\biggr)\sup_{z\in\mathsf
{X}} \biggl(\frac{\psi^{\omega}(z)}{h^{\omega}(z)} \biggr).
\end{eqnarray*}
\end{prop}

\section{Discussion}
\label{sec:Discussion}

\subsection{Sequential importance sampling}

In the case $N=1$, we have by inspection of (\ref{eq:basic_M_bold})
and (\ref{eq:G_bold}) the identity $\mathbf{M}^{\omega}\equiv M^{\omega}$,
so the particle process $ \{ \zeta_{n};n\geq0 \} $ reduces
to a Markov chain with state-space $\mathsf{X}$ and also $\mathbf
{G}^{\omega}(x)\equiv G^{\omega}(x)$.
With these observations in hand, we may apply our results to analyze
sequential importance sampling (SIS) estimators: arithmetic averages
involving independent copies of this (and other) Markov chains on~$\mathsf{X}$.

Let $\widetilde{M}\dvtx \Omega\times\mathsf{X}\times\mathcal{X}\rightarrow[0,1]$
be a Markov kernel, and for some $L\geq1$ and any fixed $\omega\in\Omega$,
let $ \{ X_{n}^{i};n\geq0 \} _{i=1}^{L}$ be $L$ i.i.d.
time-inhomogeneous Markov chains, each with law
%
\begin{equation}
X_{0}^{i}\sim\mu_{0},\qquad X_{n}^{i}|
\bigl\{ X_{n-1}^{i}=x_{n-1}^{i} \bigr\}
\sim \widetilde{M}^{\theta
^{n-1}\omega} \bigl(x_{n-1}^{i},\cdot
\bigr),\qquad n\geq1.\label{eq:SIS_law}
\end{equation}
To connect with the setting of Sections~\ref{sub:The-standard-particle}
and~\ref{sub:Alternative-sampling-of}, let $N=1$, and set $\widetilde
{\mathbf{M}}:=\widetilde{M}$.
We shall assume that {(H1)} and {(H2)} hold and that
$\widetilde{M}$ is a member of $\mathbb{M}$. With each $ \{
X_{n}^{i};n\geq0 \} $
distributed according to (\ref{eq:SIS_law}), the quantity
%
\begin{equation}
\frac{1}{L}\sum_{i=1}^{L} \Biggl[
\prod_{p=0}^{n-1}G^{\theta^{p}\omega
}
\bigl(X_{p}^{i}\bigr)\phi^{\theta^{p}\omega}
\bigl(X_{p}^{i},X_{p+1}^{i} \bigr)
\Biggr]\label{eq:SIS_estimator}
\end{equation}
is clearly an unbiased estimator of $\mathsf{E}^{\omega} [\prod_{p=0}^{n-1}G^{\theta^{p}\omega}(X_{p}^{i}) ]$.
Furthermore, since the $L$ Markov chains are independent and $\mathbf
{G}^{\omega}(x)\equiv G^{\omega}(x)$,
for any fixed $\omega$ the relative variance of (\ref{eq:SIS_estimator})
is $L^{-1}(\widetilde{\mathcal{V}}_{n,1}^{\omega}-1)$, where $\widetilde
{\mathcal{V}}_{n,1}^{\omega}$
is as in (\ref{eq:V_tilde_defn}). By application of Proposition (\ref
{prop:variance_growth})
(again with $N=1$), we have the $\mathbb{P}$-almost-sure convergence
%
\begin{equation}
\frac{1}{n}\log\widetilde{\mathcal{V}}_{n,1}^{\omega}
\longrightarrow \Upsilon_{1}(\widetilde{M}),\label{eq:SIS_var_growth}
\end{equation}
and so if $L=L(n)$,
%
\begin{equation}
\liminf_{n\rightarrow\infty}\frac{1}{n}\log \biggl(\frac
{1}{L(n)}
\widetilde{\mathcal{V}}_{n,1}^{\omega} \biggr)=\Upsilon
_{1}(\widetilde{M})-\limsup_{n\rightarrow\infty}\frac{1}{n}
\log L(n),\label{eq:SIS_lim_inf}
\end{equation}
$\mathbb{P}$-almost-surely. By Theorem~\ref{thm-variance}, except
in the case (up to the sets of measure zero mentioned therein) that
$\widetilde{M}^{\omega}(x,dx^{\prime})\propto M^{\omega}(x,dx^{\prime
})h^{\theta\omega}(x^{\prime})$,
$\Upsilon_{1}(\widetilde{M})>0$ and so the number of chains $L(n)$
must be scaled up exponentially in $n$ in order to prevent exponential
growth of the relative variance of (\ref{eq:SIS_estimator}). In this
sense the SIS approach is typically an inefficient method for approximating
$\mathsf{E}^{\omega} [\prod_{p=0}^{n-1}G^{\theta^{p}\omega
}(X_{p}^{i}) ]$,
at least relative to particle methods, which we shall now discuss.

\subsection{The bootstrap particle filter}

In the case
%
\begin{equation}
G(\omega,x):=g\bigl(x,Y_{0}(\omega)\bigr), \qquad M\bigl(
\omega,x,dx^{\prime
}\bigr):=f\bigl(x,dx^{\prime}\bigr),\label{eq:bootstrap_hmm_discuss}
\end{equation}
we have that $ \{ \mathbf{M}^{\omega};\omega\in\Omega \} $
is the collection of the transitions of the bootstrap particle filter,
as described in the \hyperref[sec:Introduction]{Introduction}. When {(H1)} and {(H2)}
hold, by Lemma~\ref{Lem:degenerate} we find that in this scenario,
for any $N\geq1$, $\Upsilon_{N}(\mathbf{M})=0$ if and only if for
$\mathbb{P}\mbox{-a.a. } \omega, \exists A_{\omega}\in\mathcal{X} \mbox
{ s.t. } \nu (A_{\omega}^{c} )=0\mbox{ and }g(x,Y_{0}(\omega
))\mbox{ is constant on }A_{\omega}$.\break
The~condition of $g(x,Y_{0}(\omega))$ being constant in $x$ represents
an entirely degenerate HMM in which the observations do not provide
any information about the hidden state. Thus, we concentrate on the
situation $\Upsilon_{N}(\mathbf{M})>0$. By an application of
Proposition~\ref{Prop:upsilon-bound}
in the case that $\psi^{\omega}(x)=1$ for all $\omega$ and $x$,
and using the bound (\ref{eq:eig_prop_bounds}) of Proposition~\ref{prop:eig},
we find that there exists a constant $c<\infty$ such that
%
\begin{equation}
\frac{1}{n}\log\widetilde{\mathcal{V}}_{n,N}^{\omega}
\rightarrow\Upsilon _{N} (\mathbf{M} )\leq\log \biggl[1+
\frac{c}{N-1} \biggr],\label
{eq:bootstrap_upsilon}
\end{equation}
where the convergence holds $\mathbb{P}$-almost surely. The practical
importance of this result is that it shows why even the rather basic
bootstrap filter is to be preferred over the SIS method in terms of
variance growth behavior, as seen by comparing~(\ref{eq:bootstrap_upsilon})
with (\ref{eq:SIS_lim_inf}).

It should be noted that under our assumptions {(H1)} and {(H2)},
the bound (\ref{eq:bootstrap_upsilon}) is implied by the bound of
\cite{smc:the:CdMG11}, Theorem~5.1. The latter also provides important
information about the nonasymptotic-in-$n$ behavior of the relative
variance, which our Proposition~\ref{Prop:upsilon-bound} does not.
On the other hand, Proposition~\ref{Prop:upsilon-bound} applies
not just to the standard particle transition $\mathbf{M}$, but also
to twisted transitions, to which the analysis of \cite{smc:the:CdMG11}
does not extend.

Continuing with the setting (\ref{eq:bootstrap_hmm_discuss}), and
assuming that {(H1)} and {(H2)} hold, we shall
now discuss $h$. The objects appearing in part (1) of Proposition~\ref{prop:eig} have the following interpretations: $\Phi_{n}^{\theta
^{-n}\omega}(\mu)\equiv\pi_{n}^{\theta^{-n}\omega}$
is the prediction filter initialized at time $-n$ using $\mu$, and
run forward to time zero, thus conditioning on the observations
$Y_{-n}(\omega),\ldots,Y_{-1}(\omega)$.
The quantity $Q_{n}^{\omega}(1)(x)$ is the conditional likelihood
of observations $Y_{0}(\omega),\ldots,Y_{n-1}(\omega)$ given that
the hidden state in the HMM at time zero is $x$. Thus, if we denote
by $\Pi_{n}^{\omega}$ the probability measure
\[
\Pi_{n}^{\omega}(A):=\frac{\int_{A}\pi_{n}^{\theta^{-n}\omega
}(dx)Q_{n}^{\omega}(1)(x)}{\int_{\mathsf{X}}\pi_{n}^{\theta^{-n}\omega
}(dz)Q_{n}^{\omega}(1)(z)},\qquad A\in\mathcal{X},
\]
we find by inspection of part (1) of Proposition~\ref{prop:eig}
that $h$ can be interpreted as the pointwise limit
%
\begin{equation}
h(\omega,x)\equiv\lim_{n\rightarrow\infty}\frac{d\Pi_{n}^{\omega}}{d\pi
_{n}^{\theta^{-n}\omega}}(x).\label{eq:bootstrap_discuss_h}
\end{equation}
Moreover, by part (2) of Proposition~\ref{prop:eig}, we find that
%
\begin{equation}
\sup_{\omega,x}\biggl\llvert \frac{d\Pi_{n}^{\omega}}{d\pi_{n}^{\theta
^{-n}\omega}}(x)-h(\omega,x)
\biggr\rrvert \leq C\rho^{n}\label
{eq:bootstrap_discuss_RN_conv}
\end{equation}
for some constants $C<\infty$ and $\rho\in(0,1)$.

Let us now consider a twisted bootstrap particle filter (as per
Section~\ref{sec:Introduction}),
in the case that for some fixed $\ell\geq1$, we take $\psi^{\omega
}:=d\Pi_{\ell}^{\omega}/d\pi_{\ell}^{\theta^{-\ell}\omega}$,
and as an instance of the setup in Section~\ref{sec:Twisted-particle-algorithms},
we let $\widetilde{\mathbf{M}}_{\ell}=\widetilde{\mathbf{M}}$ be
as per (\ref{eq:M_psi}) with this choice of $\psi^{\omega}$. We
note that $\psi^{\omega}(x)$ is proportional to the conditional likelihood,
under the HMM, of observations $Y_{0}(\omega),\ldots,Y_{\ell-1}(\omega)$
given $X_{0}=x$, and that Algorithm \ref{alg:bootstrap_twisted}
can be implemented with $\psi^{\omega}$ only specified up to a constant
of proportionality. Although typically unavailable in practice, this
$\psi^{\omega}$ allows an illustrative application of Proposition~\ref{Prop:upsilon-bound}. Indeed, using (\ref{eq:bootstrap_discuss_RN_conv}),
and the fact that under the bounds of part (3) of Proposition~\ref{prop:eig}
$h(\omega,x)$ is uniformly bounded above and below away from zero,
elementary manipulations show that there exists some finite constant
$C^{\prime}<\infty$ such that
%
\begin{equation}
\Upsilon_{N} (\widetilde{\mathbf{M}}_{\ell} )\leq\log
\biggl[1+\frac{C^{\prime}\rho^{\ell}}{N-1} \biggr].\label{eq:discuss_rho_bound}
\end{equation}
We see that, in principle, increasing the lag length $\ell$ is useful
in helping to control $\Upsilon_{N} (\mathbf{\widetilde{\mathbf
{M}}}_{\ell} )$.

Now under the mild regularity condition {(H3)}, for fixed $\omega$
and $n$, and $\varphi$ a bounded measurable function on $\mathsf{X}$,
Lemma~\ref{lem:as-convergence} shows that for the twisted particle
filter,
%
\begin{equation}
N^{-1}\sum_{i=1}^{N}\varphi\bigl(
\zeta_{n}^{i}\bigr)-\pi_{n}^{\omega}(\varphi )
\rightarrow0\label{eq:bootstrap_as}
\end{equation}
as $N\rightarrow\infty$, with probability one, independently of $\psi$.
Furthermore, by Theorem~\ref{thm:CLT}, $N^{-1/2}\sum_{i=1}^{N}
[\varphi(\zeta_{n}^{i})-\pi_{n}^{\omega}(\varphi) ]$
converges in distribution to a centered Gaussian random variable with
variance independent of $\psi$, that is, the same asymptotic variance
obtained under the standard bootstrap particle filter.

\subsection{Auxiliary particle filters}
\label{sub:Auxiliary-particle-filters}

There exist many popular alternatives to the bootstrap particle filter.
One such algorithm is the auxiliary particle filter (APF)~\cite
{pitt1999filtering},
in which current and/or future observations can influence both the
resampling and proposal of particles. In this section we consider
a family of APFs which includes the ``fully-adapted'' version of
\cite{pitt1999filtering}. Our presentation of the APF is similar
to that of \cite{johansen2008note,douc2009optimality}.

In addition to the ingredients of the HMM given in Section~\ref{sec:Introduction},
introduce $r\dvtx \Omega\times\mathsf{X}\rightarrow\mathbb{R}_{+}$ such
that for each $\omega$, $r^{\omega}(x)$ is strictly positive and
bounded in $x$. We have in mind choosing $r^{\omega}$ to be $d\Pi_{\ell
}^{\omega}/d\pi_{\ell}^{\theta^{-\ell}\omega}$
or some approximation thereof. Then set
%
\begin{eqnarray}\label{eq:apf_G_M}
G^{\omega}(x):=\frac{g(x,Y_{0}(\omega))\int_{\mathsf{X}}r^{\theta\omega
}(z)f(x,dz)}{r^{\omega}(x)},
\nonumber
\\[-8pt]
\\[-8pt]
\eqntext{ M^{\omega}\bigl(x,dx^{\prime}
\bigr)\propto f\bigl(x,dx^{\prime}\bigr)r^{\theta\omega}
\bigl(x^{\prime}\bigr).}
\end{eqnarray}
In this case, sampling according to $\mathbf{M}^{\omega}$ given by
(\ref{eq:standard_M_bold}) amounts to a form of APF. More specifically,
let $ \{ \mu_{0}^{\omega}\in\mathcal{P}(\mathsf{X});\omega\in\Omega
\} $
be the family of probability measures such that $\mu_{0}^{\omega
}(dx)\propto r^{\omega}(x)\mu_{0}(dx)$,
where $\mu_{0}$ is the initial distribution in the HMM, as in Section~\ref{sec:Introduction}. Then sampling
%
\begin{equation}
\zeta_{0}\sim \bigl(\mu_{0}^{\omega}
\bigr)^{\otimes N},\qquad \zeta _{n}|\zeta_{n-1}\sim
\mathbf{M}^{\theta^{n-1}\omega}(\zeta_{n-1},\cdot )\label{eq:apf_boldM}
\end{equation}
(we leave it to the reader to write out the algorithmic details), it
is straightforward to check using (\ref{eq:Q_GM_defn}), (\ref{eq:muQ_1=00003DE})
and (\ref{eq:lack_of_bias}) that
%
\begin{equation}
\check{Z}_{n}^{\omega,N}:=\mu_{0}\bigl(r^{\omega}
\bigr) \Biggl(\frac{1}{N}\sum_{i=1}^{N}
\frac{1}{r^{\theta^{n}\omega}(\zeta_{n}^{i})} \Biggr)\prod_{p=0}^{n-1}
\mathbf{G}^{\theta^{p}\omega}(\zeta_{p})\label{eq:apf_estimator}
\end{equation}
is an unbiased estimator of $Z_{n}^{\omega}$. If $r$ is bounded
above and below away from zero, and {(H1)} and {(H2)}
hold, then Proposition~\ref{prop:variance_growth} may be applied
to establish the existence of $\Upsilon_{N}(\mathbf{M})\geq0$ such
that the following convergence holds $\mathbb{P}$-almost surely:
%
\begin{equation}
 \frac{1}{n}\log\frac{\mathsf{E}_{N}^{\omega} [ (\check
{Z}_{n}^{\omega,N} )^{2} ]}{ (Z_{n}^{\omega}
)^{2}}\longrightarrow\Upsilon_{N}(
\mathbf{M}),\label{eq:apf_upsilon}
\end{equation}
since the $\mu_{0}(r^{\omega})$ and $N^{-1}\sum_{i=1}^{N} [r^{\theta
^{n}\omega}(\zeta_{n}^{i}) ]^{-1}$
terms have no asymptotic contribution and since the convergence in
Proposition~\ref{prop:variance_growth} is independent of the distribution
from which the particle system is initialized.

In the particular case of taking $r(\omega,x):=g(x,Y_{0}(\omega))$,
inspection of (\ref{eq:apf_G_M}) shows that we obtain the ``fully
adapted'' APF \cite{pitt1999filtering}. Moreover, Lemma~\ref{Lem:degenerate}
then shows that $\Upsilon_{N}(\mathbf{M})=0$ if and only if for $\mathbb
{P}$-almost
all $\omega$, $\int_{\mathsf{X}}g(z,Y_{1}(\omega))f(x,dz)$ is $\nu$-almost
everywhere a constant. Outside of this kind of degenerate scenario,
our analysis does not reveal whether this constant $\Upsilon_{N}(\mathbf{M})$
for the APF is bigger or smaller than the counterpart constant for
the bootstrap filter in (\ref{eq:bootstrap_upsilon}) or the constant
for the twisted bootstrap filter on the l.h.s. of (\ref{eq:discuss_rho_bound})
(with $\ell$ fixed and finite). Moreover, our analysis does not reveal
whether such orderings are invariant to the ingredients of the underlying
HMM or other elements of our setup such as the law of the observation
process, $\mathbb{P}$. Exploratory numerical experiments suggest
such invariance does not hold in general---see Section \ref
{sub:Numerical-illustrations}.

We can say something, however, about an ``ideal'' APF, arising through
a particular choice of $r$. We have seen in (\ref{eq:discuss_rho_bound})
that taking $\ell\rightarrow\infty$ in this twisted bootstrap filter,
we can push the variance growth rate to zero. There is an APF which
performs equally well in that sense; if we choose
%
\begin{equation}
r(\omega,x):=\lim_{n\rightarrow\infty}\frac{d\Pi_{n}^{\omega}}{d\pi
_{n}^{\theta^{-n}\omega}}(x),\label{eq:APF_r}
\end{equation}
that is, the generalized eigenfunction for the kernel $g(x,Y_{0}(\omega
))f(x,dx^{\prime})$,
then
%
\begin{equation}
\int_{\mathsf{X}}g\bigl(x,Y_{0}(\omega)\bigr)f
\bigl(x,dx^{\prime}\bigr)r^{\theta\omega
}\bigl(x^{\prime}\bigr)=
\chi_{\omega}r^{\omega}(x)\label{eq:APF_eigen}
\end{equation}
for a nonnegative random variable $\chi$. Applying (\ref{eq:APF_eigen})
to (\ref{eq:apf_G_M}), we find that $G^{\omega}$ appearing therein
is constant in $x$. By Lemma~\ref{Lem:degenerate}, the constant
on the r.h.s. of (\ref{eq:apf_upsilon}) then satisfies $\Upsilon
_{N}(\mathbf{M})=0$.

We can also point out a difference in how the APF and the twisted
bootstrap filter may be used to approximate integrals with respect
to the prediction filters $ \{ \pi_{n}^{\omega};n\geq0 \} $.
If for some test function $\varphi$, one wishes to use $N^{^{-1}}\sum_{i=1}^{N}\varphi(\zeta_{n}^{i})$
to approximate $\pi_{n}^{\omega}(\varphi)$ in a $N\rightarrow\infty$
consistent manner, then, in general and in contrast to (\ref{eq:bootstrap_as}),
some re-weighting must be applied to the particles. For example, assuming
(H3) holds with $G$ as in (\ref{eq:apf_G_M}) and $r^{\omega}(x)$
is bounded below away from zero in $x$, Lemma~\ref{lem:as-convergence}
and some elementary manipulations involving $\Phi_{n}^{\omega}$ show
that for bounded measurable $\varphi$,
%
\begin{equation}
\frac{\sum_{i=1}^{N}\varphi(\zeta_{n}^{i})/r^{\theta^{n}\omega}(\zeta
_{n}^{i})}{\sum_{i=1}^{N}1/r^{\theta^{n}\omega}(\zeta_{n}^{i})}-\pi _{n}^{\omega}(\varphi)
\rightarrow0\label{eq:apf_as}
\end{equation}
as $N\rightarrow\infty$, with probability $1$ under the law of the
particle system specified by (\ref{eq:apf_G_M})--(\ref{eq:apf_boldM}).
Numerical experiments (see Section~\ref{sub:Numerical-illustrations},
Figure~\ref{fig:Stochastic-Volatility-model.}) indicate that the
variance of the APF estimator in (\ref{eq:apf_as}) may be larger
than that of the bootstrap estimator (\ref{eq:bootstrap_as}). This
is perhaps attributable to the weighting of the particles in~(\ref{eq:apf_as}).

From a practical point of view, it should be noted that the computational
cost of the twisted bootstrap filter and the APF are, in general, different:
in the former, $N-1$ of the particles are propagated using the HMM
kernel $f$, whereas in the APF, all $N$ particles are propagated
using the generally more complicated kernel in (\ref{eq:apf_G_M}).
The difference in computational cost may, however, be rather dependent
on the particular model treated and the specific techniques of simulation.

Last we note that, upon assuming the setting (\ref{eq:apf_G_M})
and then following the generic structure of Section \ref
{sec:Twisted-particle-algorithms},
twisted auxiliary particle filters can readily be devised. The idea
of ``twisting'' is equally applicable to several other families
of sequential Monte Carlo algorithms.

\subsection{Numerical illustrations}
\label{sub:Numerical-illustrations}

In order to give some impression of the practical performance of the
algorithms we have analyzed, we now present some numerical findings.
{(H2)} is not satisfied for the $M$ and $G$ which specify
the models below; in this section some of our theoretical results
can only be used as guidelines for the design of practical algorithms.
We note, however, that the much milder regularity condition {(H3)}
is satisfied for the models we consider and, thus, Lemma~\ref
{lem:as-convergence}
and Theorem~\ref{thm:CLT} apply to the particle systems in question.

\begin{figure}

\includegraphics{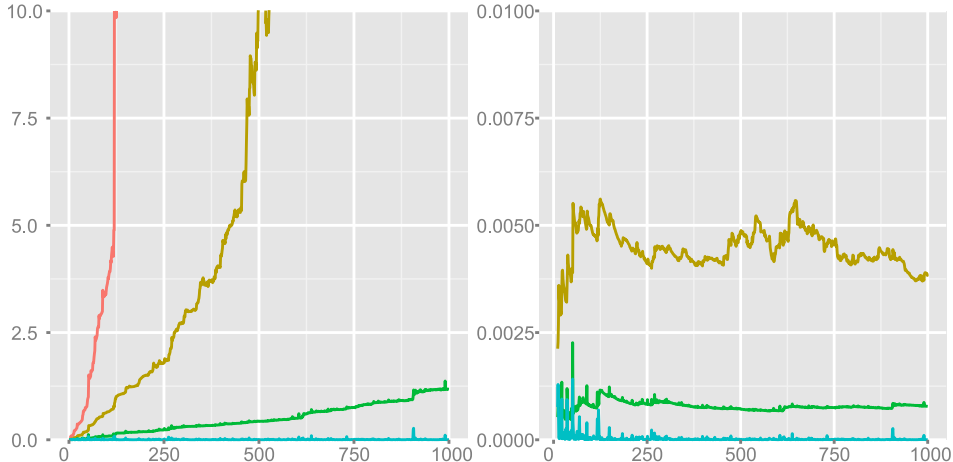}

\caption{Linear-Gaussian model and
the TPF. Estimated values of $\widetilde{\mathcal{V}}_{n,N}^{\omega}-1$
(left) and $n^{-1}\log\widetilde{\mathcal{V}}_{n,N}^{\omega}$ (right)
against $n$; with $N=100$ and $\ell=0$ (red), $\ell=1$ (yellow),
$\ell=2$ (green) and $\ell=5$ (cyan). The $\ell=0$ plot is omitted
from the right-hand figure due to scale constraints. To the precision
of these figures, increasing the lag beyond $\ell=5$ had no noticeable
influence on the variance.}\label{fig:Linear-Gaussian-model.-Left:}
\end{figure}

We shall first consider the influence of $\psi$ on the variance growth
behavior of the twisted bootstrap particle filter (henceforth TPF).
The purpose of this example is to illustrate an idealized scenario
in which $\psi^{\omega}:=d\Pi_{\ell}^{\omega}/d\pi_{\ell}^{\theta^{-\ell
}\omega}$
can be computed exactly. Consider a linear-Gaussian state-space model
where $X_{n}=0.9X_{n-1}+V_{n}$, $Y_{n}=X_{n}+W_{n}$, where $
(V_{n} ), (W_{n} )$
are i.i.d. zero mean, unit variance Gaussian sequences. Note that the
TPF algorithm can be implemented with $\psi^{\omega}$ only known
up to a constant of proportionality. Figure~\ref{fig:Linear-Gaussian-model.-Left:}
shows variance growth behavior estimated empirically using 10,000
independent runs of the algorithm for a single observation sequence,
which was drawn from the model and then fixed. Convergence of
$n^{-1}\log\widetilde{\mathcal{V}}_{n,N}^{\omega}$
is apparent and the influence of $\ell$ on the rate of variance growth
is substantial.

We now turn to a standard stochastic volatility model, in which $d\Pi
_{\ell}^{\omega}/\break d\pi_{\ell}^{\theta^{-\ell}\omega}$
is unavailable in closed form, but for which a standard deterministic
(henceforth, ``the'') approximation is available. For details of
the model, the approximation and the real data set of daily returns
on pound/dollar exchange rates, see \cite{doucet2006efficient} and
the references therein. We tested the TPF and APF using this data set
and the same model parameter settings as in the aforementioned paper.
We took both $\psi^{\omega}$ (for the TPF) and $r^{\omega}$ [for
the APF as in (\ref{eq:apf_G_M})] to both be the approximation of
$d\Pi_{\ell}^{\omega}/d\pi_{\ell}^{\theta^{-\ell}\omega}$.

\begin{figure}[t]

\includegraphics{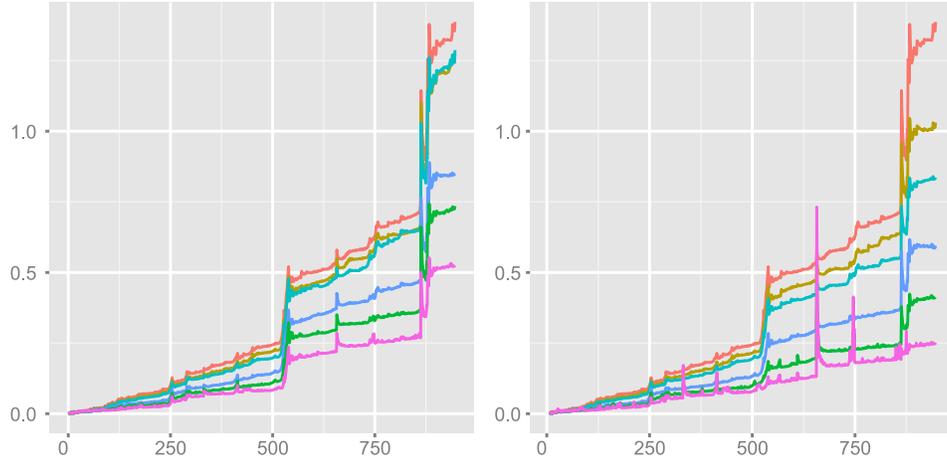}

\caption{Stochastic volatility
model. Estimated values of $\widetilde{\mathcal{V}}_{n,N}^{\omega}-1$
against $n$, for the TPF (left) and APF (right); with $N=1000$ and
$\ell=0$ (red), $\ell=1$ (yellow), $\ell=2$ (cyan), $\ell=5$ (blue)
$\ell=10$ (green) and $\ell=50$ (violet).}\label{fig:Stochastic-Volatility.-Estimated}
\end{figure}

Figure~\ref{fig:Stochastic-Volatility.-Estimated} shows empirical
variance growth behavior for a range of values of $\ell$, estimated
from 10,000 independent runs of each algorithm. For both algorithms,
increasing $\ell$ appears to generally yield a decrease in variance.
The figures indicate that, apart from occasional fluctuations, the
APF mostly exhibits lower variance than the TPF, however, we found
this phenomenon to be dependent on model parameter settings, for other
parameter values we found the TPF exhibited lower variance than the
APF (not shown).\looseness=1

\begin{figure}[b]

\includegraphics{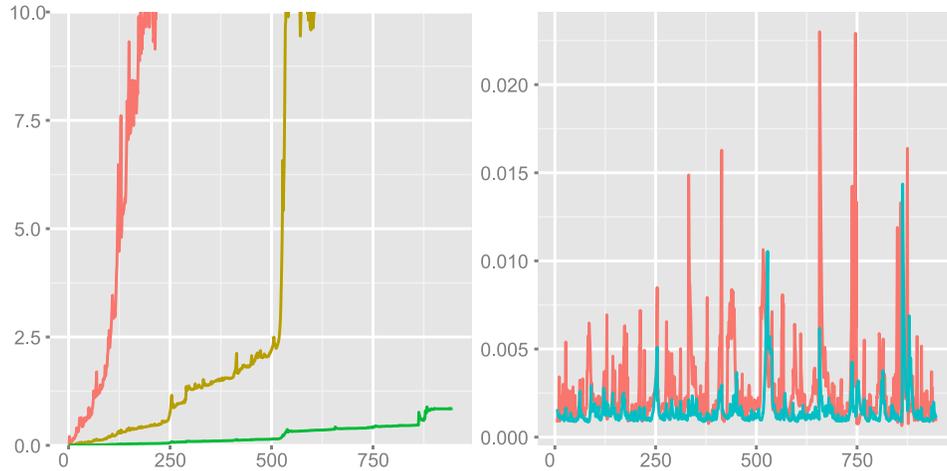}

\caption{Stochastic volatility model.
Left: estimated values of $\widetilde{\mathcal{V}}_{n,N}^{\omega}-1$
against $n$ for the TPF for $\ell=5$ and $N=10$ (red), $N=100$
(yellow) and $N=1000$ (green). Right: empirical variance of particle
approximations of the mean of $\pi_{n}^{\omega}$ against $n$, with
$N=1000$ and $\ell=5$, for the TPF (cyan) and APF (red).}\label{fig:Stochastic-Volatility-model.}
\end{figure}

The left plot in Figure~\ref{fig:Stochastic-Volatility-model.} illustrates
how the variance of the estimates from the TPF varies with $N$. An
increase in variance growth rate is evident as $N$ is decreased.
The right plot of Figure~\ref{fig:Stochastic-Volatility-model.}
shows the empirical variance of particle estimates of the mean of
the prediction filter $\pi_{n}^{\omega}$ against $n$, obtained from
the TPF and APF both with $\ell=5$. It is notable that here the TPF
generally exhibits lower variance than the APF. Results for the standard
bootstrap particle filter were found to be identical to those for
the TPF on the scale of this figure, which is in agreement with the
conclusions of Theorem~\ref{thm:CLT} applied to the TPF, that is, that
the asymptotic variance of prediction filter approximations is independent
of $\psi$.

\subsection{Generalizations and extensions}

We have only mentioned the multinomial resampling scheme, appearing
implicitly in the definition of $\mathbf{M}$ given in~(\ref
{eq:standard_M_bold}).
Several alternative schemes are popular in practice. In order to develop
extensions of Proposition~\ref{prop:variance_growth} and Theorem~\ref
{thm-variance}
to alternative schemes (assuming resampling is applied at every time
step and with a fixed number of particles), it suffices to redefine
$\mathbf{M}$ appropriately so that it incorporates the resampling
scheme of interest, to check that the conditions in the statement of
Lemma~\ref{Lem:bold_M_reg} are satisfied, and to check that Lemma~\ref
{lem:eig_bold}
holds with $\mathbf{Q}$ redefined in terms of this new $\mathbf{M}$.
Of course, the new $\mathbf{M}$ will influence the form of the corresponding
twisted algorithms.

Some types of standard particle algorithms and variants of the APF
resample according to weights which depend on two or more historical
components of the trajectory of each particle. Such algorithms can
be incorporated into the framework presented here by a simple state-space
augmentation. For example, starting from each Markov kernel $M^{\omega}$
on $(\mathsf{X},\mathcal{X})$ (according to which particles are sampled
in the algorithm of interest), one builds a kernel, $\overline
{M}^{\omega}(x,dz):=\delta_{x_{2}}(dz_{1})M^{\omega}(z_{1},dz_{2})$
on $(\mathsf{X}^{2},\mathcal{X}^{\otimes2})$, where
$x=(x_{1},x_{2}),z=(z_{1},z_{2})$
are points in $\mathsf{X}^{2}$, and introduces the appropriate incremental
weight $\overline{G}^{\omega}(x)$. Then the analyses of Section~\ref{sec:Nonnegative-kernels,-sampling}
can be repeated with mostly superficial differences: when $M^{\omega}$
satisfies (\ref{eq:H2_M}), then $\overline{M}^{\omega}$ satisfies
a 2-step version of the same condition; one then works on $(\mathsf
{X}^{2},\mathcal{X}^{\otimes2})$
instead of $(\mathsf{X},\mathcal{X})$, dealing with the kernel
$\overline{Q}^{\omega}(x,dz):=\overline{G}^{\omega}(x)\overline
{M}^{\omega}(x,dz)$
instead of~$Q^{\omega}$.

Last, we note that Proposition~\ref{prop:variance_growth} can be
easily generalized from dealing with the second moment to any $1+\delta$
moment ($\delta\geq0$), subject to suitable redefinition of $\mathbb{M}$.

\section*{Acknowledgment}

The authors thank the Associate Editor and referees for helpful comments
and suggestions.

\begin{supplement}[id=suppA]
\stitle{Twisted particle filters}
\slink[doi]{10.1214/13-AOS1167SUPP} 
\sdatatype{.pdf}
\sfilename{aos1167\_supp.pdf}
\sdescription{This supplement contains proofs of Lemmas \ref{Lem:degenerate}--\ref{lem:as-convergence},
Propositions \ref{Prop:basic-log-like}--\ref{Prop:upsilon-bound} and
Theorems~\ref{thm-variance}--\ref{thm:CLT}.}
\end{supplement}

%

%


\printaddresses

\end{document}